\documentclass[aps,prd,onecolumn,nofootinbib,groupedaddress,superscriptaddress]{revtex4}  
\usepackage{graphicx}
\usepackage{epstopdf}
\usepackage{amsmath}
\usepackage{amsfonts}
\usepackage{amssymb}
\usepackage{appendix}
\usepackage{comment}
\usepackage{bbold}
\usepackage{color}
\usepackage{slashed}
\usepackage{subfigure}
\usepackage{setspace}
\usepackage{footnote}
\usepackage{multirow}
\usepackage{longtable}
\usepackage{braket}

\newcommand{\beq}{\begin{equation}}
\newcommand{\eeq}{\end{equation}}

\begin{document}

\singlespacing

{\hfill FERMILAB-PUB-21-227-T, NUHEP-TH/21-03}

\title{$Z$-Boson Decays into Majorana or Dirac (Heavy) Neutrinos}

\author{Alain Blondel}
\affiliation{LPNHE, IN2P3/CNRS, Paris, France}
\affiliation{University of Geneva, Switzerland}
\author{Andr\'{e} de Gouv\^{e}a} 
\affiliation{Northwestern University, Department of Physics \& Astronomy, 2145 Sheridan Road, Evanston, IL 60208, USA}
\author{Boris Kayser}
\affiliation{Theoretical Physics Department, Fermilab, P.O. Box 500, Batavia, IL 60510, USA}

\begin{abstract}

We computed the kinematics of $Z$-boson decay into a heavy--light neutrino pair when the $Z$-boson is produced at rest in $e^+e^-$ collisions, including the subsequent decay of the heavy neutrino into a visible final state containing a charged-lepton. We concentrated on heavy-neutrino masses of order dozens of GeV and the issue of addressing the nature of the neutrinos -- Dirac fermions or Majorana fermions. We find that while it is not possible to tell the nature of the heavy and light neutrinos on an event-by-event basis, the nature of the neutrinos can nonetheless be inferred given a large-enough sample of heavy--light neutrino pairs. We identify two observables sensitive to the nature of neutrinos. One is the forward-backward asymmetry of the daughter-charged-leptons. This asymmetry is exactly zero if the neutrinos are Majorana fermions and is non-zero (and opposite) for positively- and negatively-charged daughter-leptons if the neutrinos are Dirac fermions. The other observable is the polarization of the heavy neutrino, imprinted in the laboratory-frame energy distribution of the daughter-charged-leptons. Dirac neutrinos and antineutrinos produced in $e^+e^-$ collisions at the $Z$-pole are strongly polarized while Majorana neutrinos are at most as polarized as the $Z$-bosons. 

\end{abstract}


\maketitle

\section{Introduction}
\label{sec:intro}

Spin one-half massive fermions in four spacetime dimensions can be classified into two types: Majorana fermions and Dirac fermions. Majorana fermions are self-conjugate; they are their own antiparticles. Dirac fermions are not. A massive Dirac field describes four independent degrees of freedom, a massive Majorana field describes two independent degrees of freedom. Dirac fields are complex and can transform under complex representations of symmetry groups (gauged or global). Majorana fields are, in some sense, real and can only transform under real representations of unbroken symmetries (gauged or global). 

All known fundamental fermions are massive, except, at most, the lightest of the three neutrinos. With the exception of the neutrinos, all have nonzero electric charge. Hence, the charged leptons ($e,\mu,\tau$) and all quarks are Dirac fermions. Experimental data to date are agnostic concerning the nature of the neutrinos -- Majorana or Dirac fermions. 

The traditional way to probe the nature of neutrinos is to test the conservation of lepton number (see, for example, \cite{deGouvea:2013zba,Dolinski:2019nrj}). Majorana neutrinos, since they are self-conjugate, cannot be charged under the lepton-number symmetry, rendering it, necessarily, an approximate symmetry. Evidence for the violation of lepton-number symmetry (by two units) would translate, almost necessarily, into proof that neutrinos are Majorana fermions \cite{Schechter:1981bd,Nieves:1984sn,Takasugi:1984xr} (see also \cite{Hirsch:2006yk} for a modern discussion). The most powerful probes of lepton-number violation are low-energy particle and nuclear physics processes, including searches for neutrinoless double-beta decay, $(\mu^-\to e^+)$-conversion in nuclei, and forbidden meson decays (e.g. $K^+\to\pi^-\mu^+\mu^+$). No unambiguous experimental evidence for lepton-number violation has been found yet \cite{Zyla:2020zbs}. 

The physics behind nonzero neutrino masses remains unknown. One can, however, affirm that nonzero neutrino masses imply the existence of new particles (for an overview see, for example, \cite{Gouvea:2016shl}). The properties -- spin, mass, charge --  of these new particles is currently unknown and the subject of extensive theoretical and experimental particle physics research. Information on the nature of neutrinos is expected to illuminate the origin of neutrino masses in a very fundamental way.

New neutral fermions are common side effects of the mechanism responsible for nonzero neutrino masses. The popular Type-I seesaw mechanism \cite{Minkowski:1977sc,GellMann:1980vs,Yanagida:1979as,Glashow:1979nm,Mohapatra:1979ia,Schechter:1980gr}, for example, posits the existence of at least two gauge-singlet fermion fields in addition to the known matter fields of the Standard Model. These are allowed nonzero Majorana-mass terms and couple to the active neutrinos $\nu_e,\nu_{\mu},\nu_{\tau}$ via Yukawa interactions. After electroweak symmetry breaking, all neutral fermions -- linear combinations of the active neutrinos and the new degrees of freedom  -- acquire masses, all are Majorana fermions, and all couple with different strength to the electroweak gauge bosons and the Higgs boson. In the traditional Type-I seesaw, the neutral fermions ``split'' into three mostly-active neutrinos that are very light and at least two much heavier states that are predominantly composed of gauge-singlet fermions. We refer to the former as light neutrinos and the latter as heavy neutrinos. In this scenario, the couplings of the heavy neutrinos are related to the light and heavy neutrino masses. Searches for heavy neutrinos, therefore, allow one to verify whether neutrino masses have anything to do with the Type-I seesaw mechanism. The discovery of heavy neutrinos, of course, would not imply that neutrino masses are a consequence of the Type-I seesaw mechanism or one of its variations. Progress on that front will require the measurement of the properties of the heavy neutrinos, including their nature. The Type-I seesaw, for example, predicts that the light and heavy neutrinos are Majorana fermions. There are, however, perfectly reasonable new-physics models that posit the existence of heavy Dirac neutrinos, including scenarios inspired by the dark matter puzzle (see, for example, \cite{deGouvea:2015pea}) and those aimed at explaining the small Dirac neutrino-masses (see, for example, \cite{ArkaniHamed:1998vp,Dienes:1998sb,Dvali:1999cn}). 

Heavy neutrinos produced at colliders allow other searches for lepton-number violation. For example, in a high-energy $pp$ collider, heavy neutrinos can be produced via charged-current processes and decay, for example, into charged leptons and $W$-bosons (on or off-shell). We provide more details in Sec.~\ref{sec:heavynu} but here it suffices to state that a heavy Dirac neutrino $\nu_4$ carries lepton-number and must decay to a final state with lepton number $+1$, e.g., $\nu_4\to\ell^-W^{(*)+}$. The heavy antineutrino $\bar{\nu}_4$ will decay into a final state with lepton number $-1$, e.g., $\bar{\nu}_4\to\ell^+W^{(*)-}$. The Majorana heavy neutrino, instead, can decay into positively and negatively charged-leptons with equal probability at leading order. Hence, if the heavy neutrinos are Majorana fermions, the following process, and others like it, should occur:
\begin{equation}
pp\to \ell^++\nu_4+X \to \ell^++(\ell^+W^{(*)-})+X,
\label{eq:LHC}
\end{equation}
where $X$ is a state with zero lepton number (e.g., hadrons). As long as the lepton number of $X$ is known to be zero,\footnote{For example, when there are no light neutrinos in $X$. These would manifest themselves as missing transverse energy at collider experiments like those at the LHC.} Eq.~(\ref{eq:LHC}) is a lepton-number violating process. Its observation would reveal that the heavy neutrinos are Majorana fermions. 

Other than the observation of explicit lepton-number violation, there are several other ways through which Dirac and Majorana fermions can be distinguished. In \cite{Balantekin:2018ukw} (see also \cite{Balantekin:2018azf}), a subset of the authors of this manuscript highlighted that Dirac and Majorana fermions decay ``differently'' (see \cite{Shrock:1982sc,Li:1981um} for early discussions). In more detail, it was demonstrated that when a Majorana fermion decays at rest into a Majorana daughter and a self-conjugate boson (e.g., $Z$, $h$, $\pi^0$, or $\gamma$), the angular distribution of the daughters is isotropic, regardless of the polarization of the parent fermion. Instead, if the parent and daughter fermions are Dirac fermions, the angular distribution of the final state can be nontrivial. Hence, the observation of a nontrivial angular distribution of $\nu_4$ decays into, for example, $\nu+\gamma$, would imply, necessarily, that $\nu_4$ are Dirac fermions. 

Here, we study the production of heavy neutrinos in the decay of on-shell $Z$-bosons produced at rest via electron-positron collisions. In the last several years, electroweak scale circular electron-positron colliders have been proposed as `Higgs Factories:' FCC-ee~\cite{Benedikt:2651299} and CEPC~\cite{CEPCStudyGroup:2018rmc,CEPCStudyGroup:2018ghi}. They offer the possibility of a very high luminosity run at the $Z$-pole and the ability to observe over $10^{12}$ $Z$-boson decays. Given current constraints on heavy-neutrino production, there is the possibility to produce and identify thousands of heavy neutrinos with masses of order tens of GeV. The experimental conditions are particularly favorable since there is no pile-up in these machines and the heavy neutrinos are produced in the simple two-body decay of a heavy particle at rest. The question of observability of the process will be quickly reviewed in Sec.~\ref{sec:heavynu}. Immediately after the discovery of a heavy neutrino candidate from the observation of a few events, the question would be raised: is this particle a Majorana-fermion or a Dirac-fermion?
%
%
%
We find that it is possible to determine the nature of the heavy neutrinos with lepton-collider data on the $Z$-pole. This is in spite of the fact that total lepton number cannot be measured on an event-by-event basis and hence one is blind to explicit lepton-number violating effects. 

As already advertised, we define heavy neutrinos in Sec.~\ref{sec:heavynu}, along with other useful information. We discuss $Z$-boson decays into a heavy--light neutrino pair in Sec.~\ref{sec:Zdecay} and show, in Sec.~\ref{sec:Ndecay}, how the decays of the heavy neutrinos might be used, when these are produced in $Z$-boson decays, to determine their nature. Some concluding remarks follow in Sec~\ref{sec:conclusion}.

%

\section{Heavy Neutrinos}
\label{sec:heavynu}
\setcounter{equation}{0}

Heavy neutrinos are massive neutral fermions that couple to the electroweak gauge bosons via mixing with the active neutrinos. In other words, there are $n\ge 3$ neutrinos $\nu_i$ with masses $m_i$ ($i=1,2,3,\ldots,n$ ) and three linear combinations of those couple to the charged-leptons and the $W$-boson:
\begin{equation}
-{\cal L} \supset \sum_{\alpha=e,\mu,\tau}\sum_{i=1}^n \frac{g}{\sqrt{2}} \left(U_{\alpha i}\bar{\ell}_{\alpha}\slashed{W}^-P_L\nu_i\right) + H.c.~,
\end{equation}
where $P_L=(1-\gamma_5)/2$ is the left-chirality-projection operator, $g$ is the $SU(2)$ coupling constant, and $U_{\alpha i}$ are the elements of a unitary $n\times n$ matrix. We indicate the summations over $\alpha$ and $i$ explicitly to highlight there are  $n$ neutrinos but only three charged-leptons. Similarly, the couplings of $\nu_i$ and $\ell_{\alpha}$ to the $Z$-boson are
\begin{equation}
-{\cal L} \supset \frac{g}{2\cos\theta_W}\left[  \sum_{\alpha=e,\mu,\tau}\sum_{i,j=1}^n \left(U_{\alpha i}U^*_{\alpha j}\bar{\nu}_{j}\slashed{Z}P_L\nu_i\right) + \sum_{\alpha=e,\mu,\tau} \left(g_L \bar{\ell}_{\alpha}\slashed{Z}P_L\ell_{\alpha}+ g_R \bar{\ell}_{\alpha}\slashed{Z}P_R\ell_{\alpha}\right)\right]~,
\end{equation}
where $P_R=(1+\gamma_5)/2$ is the right-chirality-projection operator and $g_L=(1-2\sin^2\theta_W)$  and $g_R=2\sin^2\theta_W$ are the left- and right-chiral charges of the charged-leptons to the $Z$-boson, respectively. As usual, $\theta_W$ is the weak mixing angle (Weinberg angle).

If $n=3$, because $U$ is a unitariy matrix, $\sum_{\alpha=e,\mu,\tau} U_{\alpha i}U^*_{\alpha j}=\delta_{ij}$ and, as is well known, the neutrino--$Z$-boson couplings are diagonal in the mass basis. This is not the case if $n>3$. The coupling between $\nu_4$ and $\nu_1$, for example, is proportional to $\sum_{\alpha}U_{\alpha_1}U^*_{\alpha 4}$. This does not vanish since the $\alpha$ sum is restricted to $e,\mu,\tau$. 

We are interested in the decays of the $Z$-boson into a heavy neutrino and a light neutrino.\footnote{If the neutrinos are Dirac fermions, henceforth `neutrino' is meant to stand for both the neutrino and the antineutrino whenever it does not lead to unnecessary confusion.} We define the mass eigenvalues in such a way that $m_{1,2,3}$ are the light neutrino masses and $m_{4,5,\ldots}$ the heavy ones. The partial width of the Z-boson to decay into a $\nu_4$ and any light neutrino is 
\begin{equation}
\Gamma(Z\to \nu_4\nu_{\rm light})\propto \sum_{\alpha,\beta=e,\mu,\tau}\sum_{i=1,2,3}U_{\alpha i}U^*_{\beta i} U^*_{\alpha 4}U_{\beta 4} ~\sim \sum_{\alpha=e,\mu,\tau} |U_{\alpha 4}|^2\equiv |U_4|^2~,
\end{equation}
where we take advantage of the fact that we are interested in the limit where the $3\times 3$ ``light'' leptonic mixing submatrix is very close to a unitary matrix. The partial width into two heavy neutrinos, in this limit, is negligibly small. For example, $\Gamma(Z\to \nu_4\nu_4)\propto |U_4|^4$. More quantitatively \cite{Dittmar:1989yg}, under the assumption that the $3\times 3$ ``light'' leptonic mixing submatrix is very close to a unitary matrix,
\begin{equation}
B(Z\to \nu_4\nu_{\rm light}) = 2|U_4|^2 \frac{B(Z\to{\rm invisible})}{3}\left(1+\frac{m_4^2}{2M_Z^2}\right)\left(1-\frac{m_4^2}{M_Z^2}\right)^2,
\end{equation}
where $B(Z\to{\rm invisible})\simeq 20\%$ is the invisible $Z$-boson branching ratio and $M_Z$ is the $Z$-boson mass. For $m_4=10$~GeV, $B(Z\to \nu_4\nu_{\rm light})\simeq0.13|U_4|^2$ while $B(Z\to \nu_4\nu_{\rm light})\simeq0.0104|U_4|^2$ for $m_4=79$~GeV. Hence, at least one heavy neutrino would be expected from $10^{12}$ $Z$-boson decays (Tera-$Z$) if $|U_{4}|^2>7.5\times 10^{-12}$ for $m_4=10$~GeV (or $|U_{4}|^2>9\times 10^{-11}$ for $m_4=79$~GeV).

The heavy neutrino is potentially observable if it decays visibly inside the detector-volume. The decay of heavy neutrinos is mediated by $W$-boson and $Z$-boson exchange. The partial decay-width of the $\nu_4$ is proportional to $|U_{\alpha 4}|^2$ for the decay into an $\ell_{\alpha}$. Summing over all final-state charged leptons, assuming $m_4\gg m_{\tau}$, the charged-lepton partial width is proportional to $|U_{4}|^2$. The partial decay-width into light neutrinos is also proportional to $|U_4|^2$, including all final-state light neutrinos. For heavy neutrino masses above 10~GeV but below the $W$-boson mass, we roughly estimate, taking into account that, for the $m_4$ values of interest, the charged-current decay mode accounts for the majority of the decays \cite{Dittmar:1989yg},
\begin{equation}
c\tau_4 \sim \frac{10^{-6}}{|U_4|^2}\left(\frac{10~{\rm GeV}}{m_4}\right)^5 {\rm cm}~.
\end{equation}

The strongest bounds on the production of heavy neutrinos in $e^+e^-$-scattering followed by visible $\nu_4$ decays come from the DELPHI experiment at LEP \cite{Abreu:1996pa}. They constrain $|U_{4}|^2$ to be less than about $10^{-5}$ for masses between roughly 5~GeV and 70~GeV. For smaller masses, the decay is too slow and the heavy neutrinos decay outside the detector while for larger masses the phase-space suppression is too severe. This proves to be the strongest bound on heavy neutrinos in this mass range, see, for example, \cite{Drewes:2015iva,Fernandez-Martinez:2016lgt,Drewes:2016jae,Bryman:2019ssi,Bryman:2019bjg,Bolton:2019pcu}. Recent studies of the sensitivity of future $e^+e^-$ ``$Z$-factories'' to the production of heavy neutrinos can be found in, for example, \cite{Blondel:2014bra,Antusch:2016vyf,Liao:2017jiz,Blondel:2018mad,Ding:2019tqq}. They demonstrate that next-generation $e^+e^-$ collider could be sensitive to  $|U_{4}|^2\gtrsim 10^{-11}$ in the $m_4\sim[5,70]$~GeV mass window. 

\section{$Z$-Boson Decay to a Heavy-Light Neutrino Pair}
\label{sec:Zdecay}
\setcounter{equation}{0}

We are interested in the processes $e^+e^-\to Z\to \nu_4\bar{\nu}_i$ and $e^+e^-\to Z\to \nu_i\bar{\nu}_4$, if the neutrinos are Dirac fermions, or the process  $e^+e^-\to Z\to \nu_4\nu_i$ if they are Majorana fermions. As mentioned in the previous section, we will compute the differential cross-section summing over all light neutrinos $\nu_i$, $i=1,2,3$. We assume the collision happens at the $Z$-pole and that the $Z$-bosons are produced at rest. In our calculations, we ignore the $t$-channel $W$-exchange contribution to neutrino production and assume the electron and the light neutrinos are massless. These are very safe assumptions. We allow, however, for nonzero heavy neutrino masses $m_4$.

If we define the $z$-axis to coincide with the direction of the electron three-momentum, the $Z$-bosons are produced in eigenstates of $S_z$ with eigenvalues $\pm1$. The $S_z=0$ fraction of the population is proportional to the electron mass, hence negligible. Parity violation translates into a slight asymmetry and the $Z$-boson population associated to $S_z=-1$ is slightly larger. In this sense, we say the $Z$-boson sample is polarized. Quantitatively, the $Z$-boson polarization is 
\begin{equation}
P_Z = \frac{(g_R^2-g_L^2)}{(g_L^2+g_R^2)}.
\label{eq:PZ}
\end{equation}
Numerically, $P_Z\simeq -0.15$.

\begin{figure}[ht]
\includegraphics[width=0.49\textwidth]{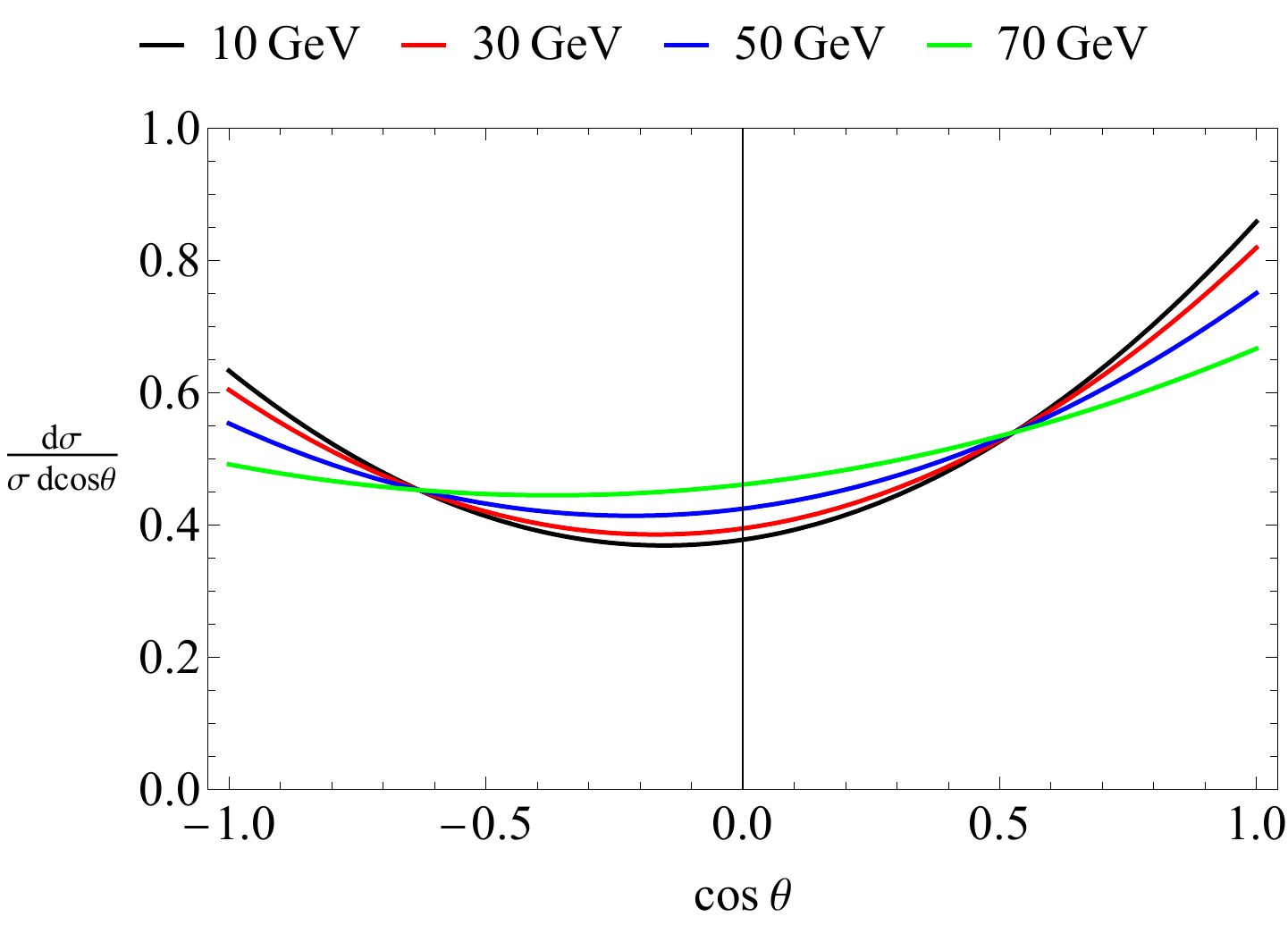}
\includegraphics[width=0.49\textwidth]{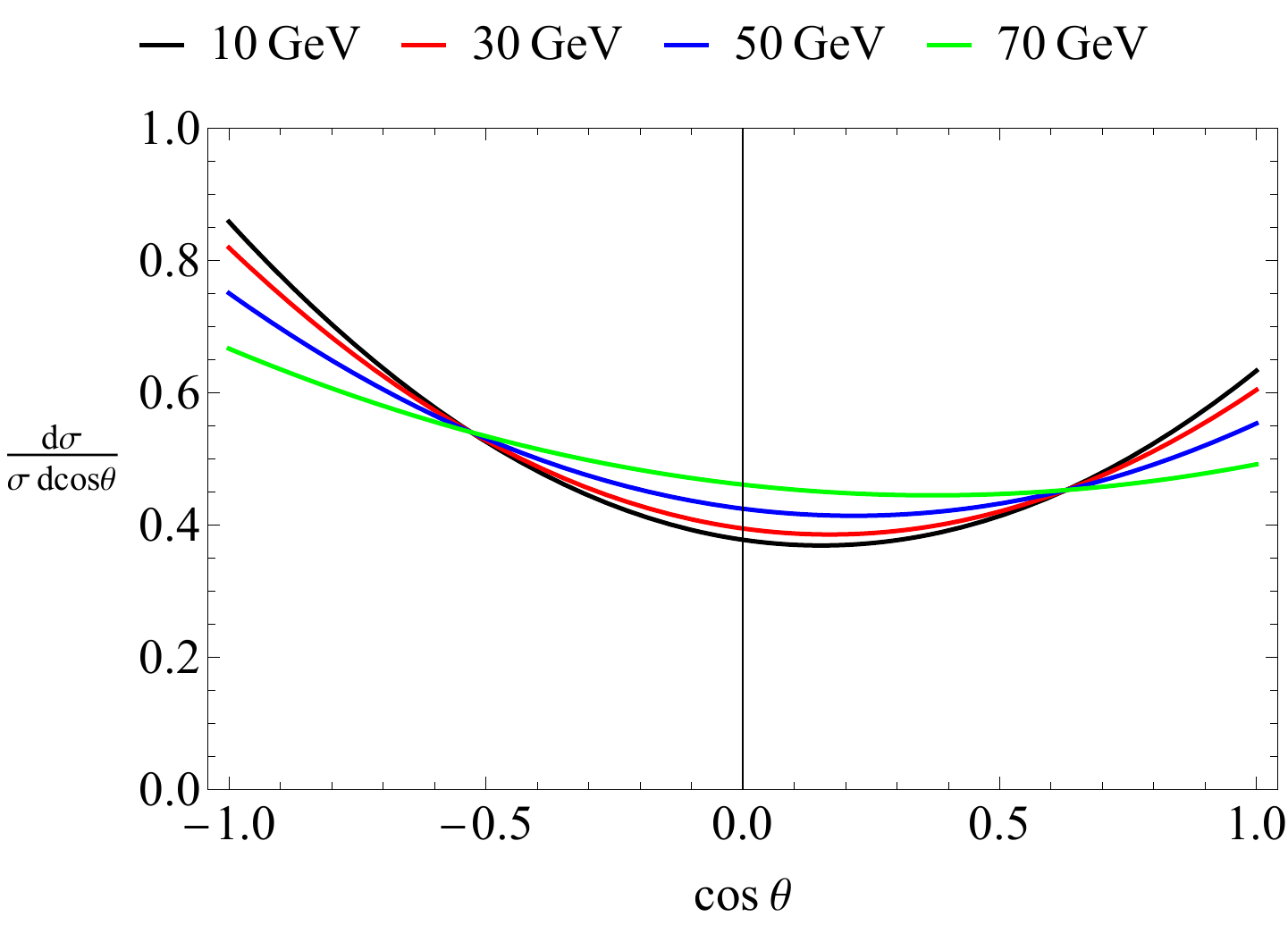}
\caption{Normalized differential cross-section for $e^+e^-\to Z \to \nu_4\bar{\nu}_{\rm light}$ (left) and $e^+e^-\to Z \to \bar{\nu}_4\nu_{\rm light}$ (right) as a function of the direction of the heavy (anti)neutrino $\cos\theta$, for different values of the heavy neutrino mass $m_4$. The neutrinos are assumed to be Dirac fermions.} 
\label{fig:DistD}
\end{figure}

The parity-violating nature of the weak interactions implies a potential forward-backward asymmetry of the $\nu_i$ and the $\nu_4$ (and their antiparticles, if applicable) and nontrivial polarization. If the neutrinos are Dirac fermions, we find
\begin{eqnarray}
\frac{1}{\sigma_D(\nu_4)}\frac{{\rm d}\sigma_D(\nu_4)}{{\rm d}\cos\theta} = \frac{3}{4(g_R^2+g_L^2)}\frac{M_Z^2}{(2M_Z^2+m_4^2)}\left(g_R^2(1-\cos\theta)^2+g_L^2(1+\cos\theta)^2+ \frac{m_4^2}{M_Z^2}(g_R^2+g_L^2)\sin^2\theta
\right)~, \\
\frac{1}{\sigma_D(\bar{\nu}_4)}\frac{{\rm d}\sigma_D(\bar{\nu}_4)}{{\rm d}\cos\theta} = \frac{3}{4(g_R^2+g_L^2)}\frac{M_Z^2}{(2M_Z^2+m_4^2)}\left(g_R^2(1+\cos\theta)^2+g_L^2(1-\cos\theta)^2+ \frac{m_4^2}{M_Z^2}(g_R^2+g_L^2)\sin^2\theta
\right)~,
\end{eqnarray}
where $\sigma_D(\nu_4)$, and $\sigma_D(\bar{\nu}_4)$ are the cross-sections for $e^+e^-\to Z\to \nu_4\bar{\nu}_i$ and $e^+e^-\to Z\to \nu_i\bar{\nu}_4$, respectively, assuming the neutrinos are Dirac fermions. $\theta$ is the angle the heavy neutrino three-momentum defines relative to the direction of the electron three-momentum. The term proportional to $(m_4/M_Z)^2$ is the ``wrong-helicity'' contribution, where the light neutrino and heavy antineutrino (or vice-versa) are emitted with the same helicity. This contribution is invariant under $\cos\theta\to-\cos\theta$ and vanishes when $\cos\theta=\pm1$. This can be understood in terms of naive angular momentum conservation. When the daughter neutrinos are emitted in the $z$-direction, they are eigenstates of $S_z$ with eigenvalue $\pm1/2$. Since they are back-to-back, their helicities must be opposite to one another. The angular distribution of the $\nu_4$ and $\bar{\nu}_4$ are depicted in Fig.~\ref{fig:DistD}, for different values of $m_4$.

\begin{figure}[ht]
\includegraphics[width=0.55\textwidth]{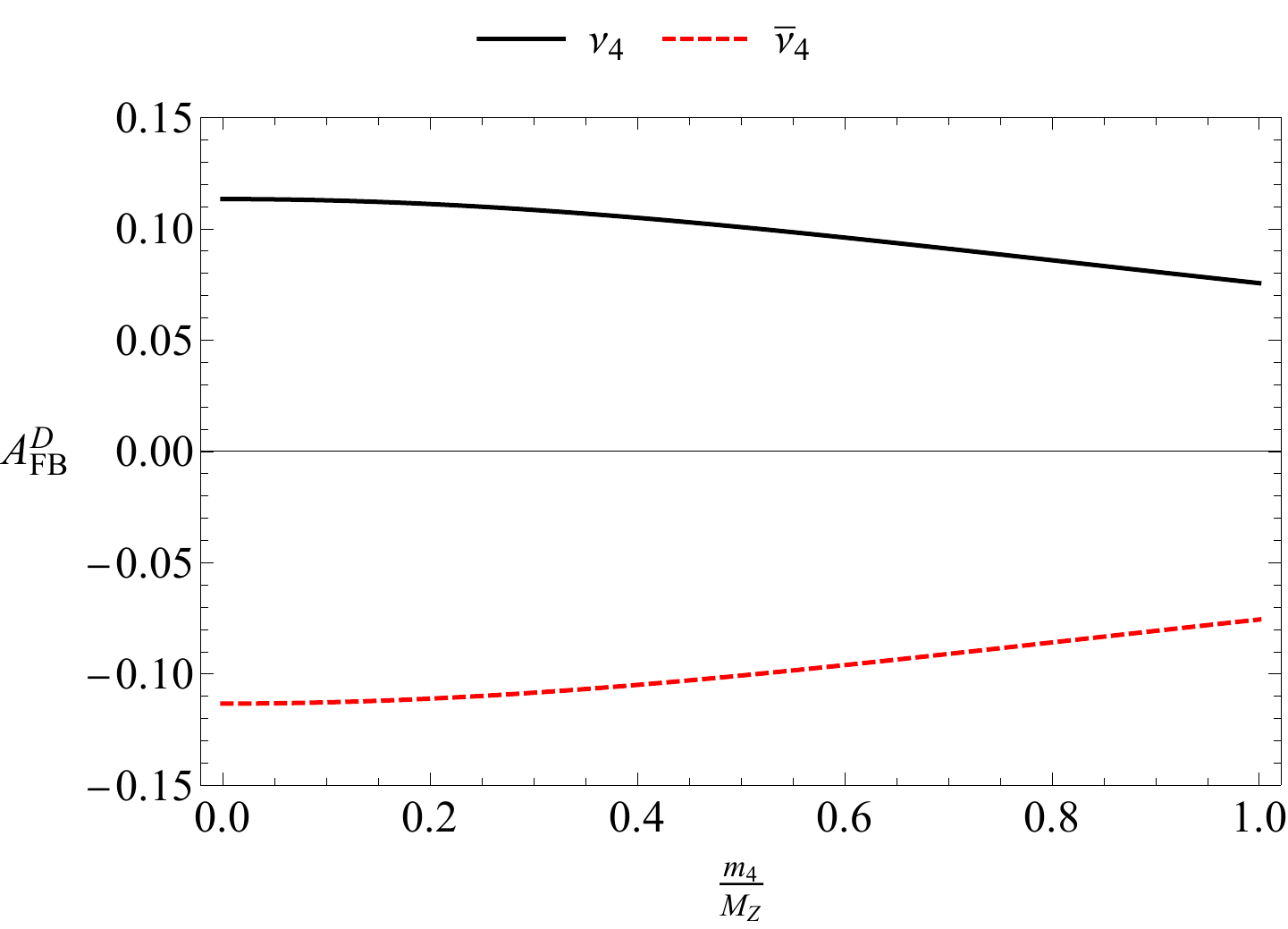}
\caption{The Forward-Backward Asymmetry $A^D_{FB}$ of heavy neutrino or antineutrino production in $e^+e^-\to Z \to \nu_4\bar{\nu}_{\rm light}$ or $e^+e^-\to Z \to \bar{\nu}_4\nu_{\rm light}$ as a function of the heavy neutrino mass. The neutrinos are assumed to be Dirac fermions.} 
\label{fig:AFB_D}
\end{figure}

Parity violation ($g_L^2\neq g_R^2$) implies a forward-backward asymmetry for Dirac-fermion $\nu_4$ and $\bar{\nu}_4$ production. As usual, we define
\begin{equation}
A_{\rm FB} = \frac{1}{\sigma}\left[{\int_{0}^1\frac{{\rm d}\sigma}{{\rm d}\cos\theta}}{\rm d}\cos\theta-{\int_{-1}^0\frac{{\rm d}\sigma}{{\rm d}\cos\theta}}{\rm d}\cos\theta\right].
\end{equation}
For Dirac-fermion $\nu_4$ and $\bar{\nu}_4$ production we find, respectively,
\begin{eqnarray}
A^D_{\rm FB}(\nu_4)= \frac{3}{2} \frac{M_Z^2}{(2M_Z^2+m_4^2)} \frac{(g_L^2-g_R^2)}{(g_L^2+g_R^2)}~, \label{eq:AFBD} \\
A^D_{\rm FB}(\bar{\nu}_4)= \frac{3}{2} \frac{M_Z^2}{(2M_Z^2+m_4^2)} \frac{(g_R^2-g_L^2)}{(g_L^2+g_R^2)}~. \label{eq:antiAFBD}
\end{eqnarray}
The forward-backward  asymmetries for $\nu_4$ and $\bar{\nu}_4$ production are depicted in Fig.~\ref{fig:AFB_D} as a function of $m_4$.

\begin{figure}[ht]
\includegraphics[width=0.60\textwidth]{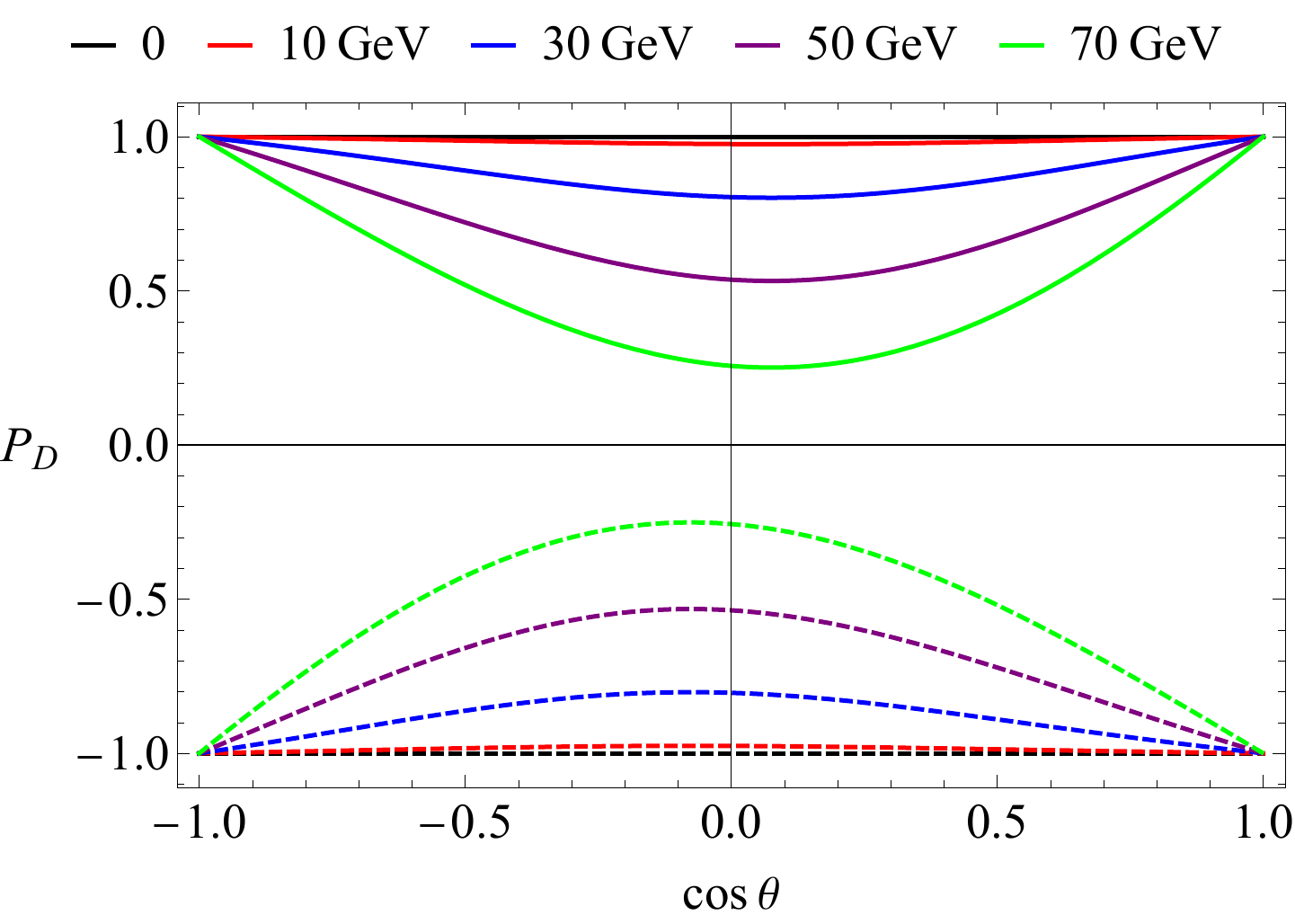}
\caption{The polarizarion  $P_D$ of heavy neutrinos (dashed lines) or antineutrinos (solid lines) produced in $e^+e^-\to Z \to \nu_4\bar{\nu}_{\rm light}$ or $e^+e^-\to Z \to \bar{\nu}_4\nu_{\rm light}$ as a function of the direction of the heavy (anti)neutrino $\cos\theta$, for different values of the heavy neutrino mass $m_4$. The neutrinos are assumed to be Dirac fermions.} 
\label{fig:PD}
\end{figure}

We are also interested in the polarization of the heavy neutrinos. That of the light neutrinos is simple: if the neutrinos are Dirac fermions, all light neutrinos are left-handed -- negative helicity -- and all light antineutrinos are right-handed -- positive helicity. We computed the polarization of the heavy neutrino $P(\cos\theta)$ as a function of the production angle $\theta$ of $\nu_4$ and $\bar{\nu}_4$. The results are straight-forward but a little cumbersome so we don't write them out explicitly here. $P(\cos\theta)$ is depicted in Fig.~\ref{fig:PD} for $\nu_4$ and $\bar{\nu}_4$, for different values of $m_4$. $P=1$ corresponds to a positive-helicity state (right-handed) while $P=-1$ corresponds to a negative-helicity state (left-handed).

Many of the features of Fig.~\ref{fig:PD} are easy to understand. For small heavy neutrino masses, $m_4\ll M_Z$, the helicity of the heavy neutrino is dictated by the left-chiral nature of the neutrino--$Z$-boson coupling. In this limit, all heavy neutrinos are left-handed ($P=-1$) while all heavy antineutrinos are right-handed ($P=+1$).  For $\cos\theta=\pm1$, as already argued, the helicity of the heavy neutrino and light antineutrino (and vice-versa) are equal and opposite because of angular-momentum conservation. Since the light neutrinos and antineutrinos are always 100\% polarized, for $\cos\theta=\pm1$, so are the heavy neutrinos, independent from $m_4$.  For finite $m_4$, the magnitude of the polarization decreases as $\cos^2\theta$ decreases. This is a reflection of the fact that the ``wrong-helicity'' distribution peaks at $\cos^2\theta\to 0$. Finally, the forward-backward asymmetry of $P(\cos\theta)$ is a consequence of the nontrivial polarization of the decaying $Z$-boson.

\begin{figure}[ht]
\includegraphics[width=0.55\textwidth]{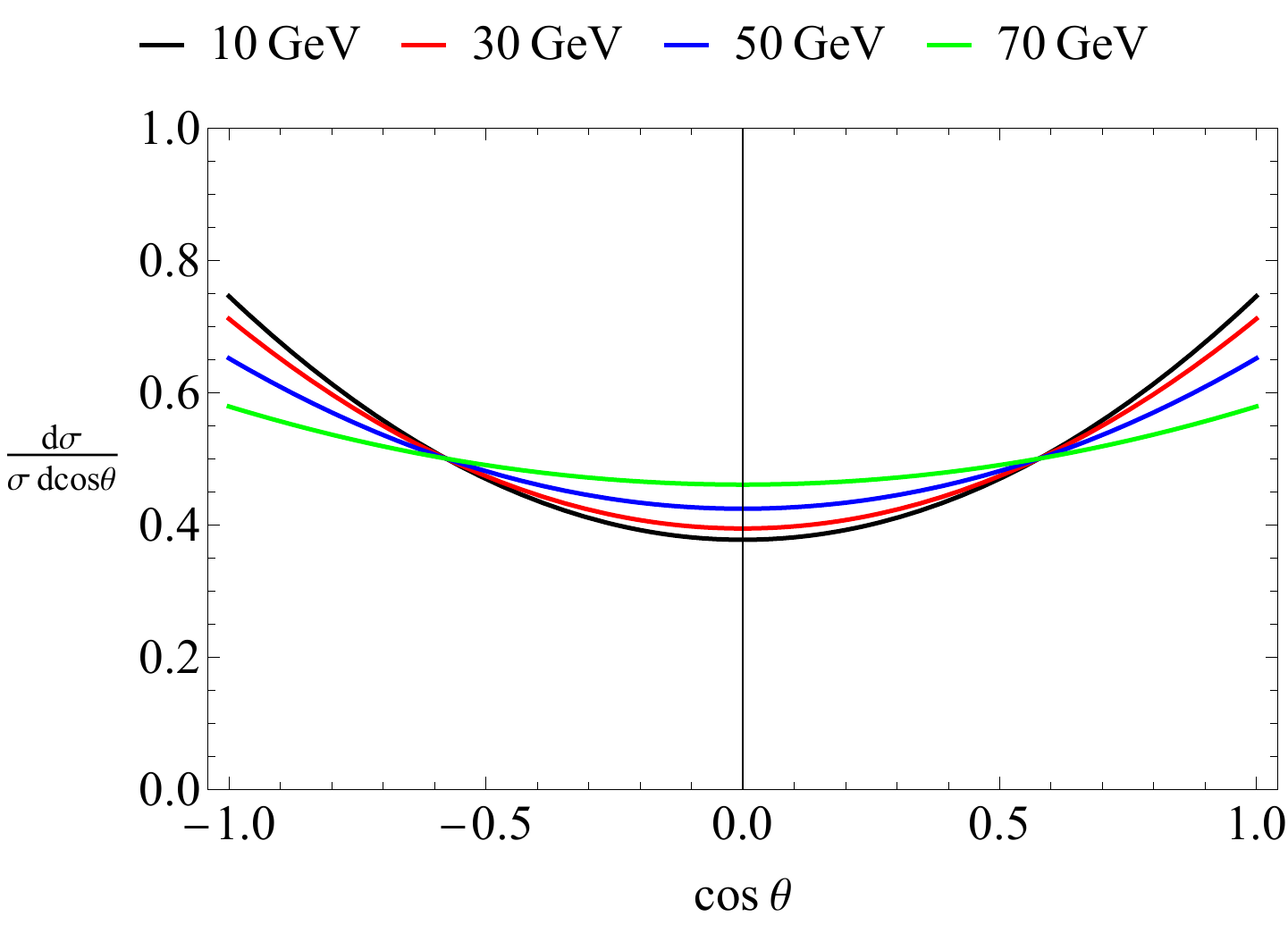}
\caption{Normalized differential cross-section for $e^+e^-\to Z \to \nu_4\nu_{\rm light}$ as a function of the direction of the heavy neutrino, $\cos\theta$, for different values of the heavy neutrino mass $m_4$. The neutrinos are assumed to be Majorana fermions.} 
\label{fig:DistM}
\end{figure}

If the neutrinos are Majorana fermions, there is only one process of interest: $e^+e^-\to Z\to \nu_4\nu_i$, $i=1,2,3$. In this case, we find,
\begin{equation}
\frac{1}{\sigma_M(\nu_4)}\frac{{\rm d}\sigma_M(\nu_4)}{{\rm d}\cos\theta} = \frac{3}{4}\frac{M_Z^2}{(2M_Z^2+m_4^2)}\left(1+\cos^2\theta +\frac{m_4^2}{M_Z^2}\sin^2\theta
\right)~,
\end{equation}
where $\sigma_M(\nu_4)$ is the cross-section for $e^+e^-\to Z\to \nu_4\nu_i$ assuming the neutrinos are Majorana fermions. This distribution is invariant under $\cos\theta\to-\cos\theta$ so the forward-backward asymmetry vanishes exactly: $A_{\rm FB}^M(\nu_4)=0$. The angular distribution of the Majorana $\nu_4$ is depicted in Fig~\ref{fig:DistM}, for different values of $m_4$.

The absence of a forward-backward asymmetry in the decay of a self-conjugate vector boson ($Z$-boson) into two self-conjugate fermions ($\nu_4\nu_{\rm light}$) is a general leading-order consequence of the CPT-theorem. Our proof and notation parallel those at other recent discussions \cite{Balantekin:2018ukw,deGouvea:2021ual}. At leading order in perturbation theory, the amplitude for the decay of a $Z$-boson at rest in an eigenstante of $S_Z$ with eigenvalue $+1$ (denoted by `$\Uparrow$') into a $\nu_4$ with helicity $\lambda_4$ emitted in a direction characterized by the angle $\theta$ and a light neutrino characterized by helicity $\lambda_\nu$ is 
\begin{equation}
\langle \nu_4(\theta, \lambda_4)\, \nu_{\rm light}(\pi-\theta,\lambda_\nu) \;|\; {\cal H}_{\mathrm{int}}\; |\; Z(\mathrm{\Uparrow}) \rangle \;\; ,
\label{eq:Hdecay}
\end{equation}
where ${\cal H}_{\mathrm{int}}$ is the interaction Hamiltonian. We assume  that ${\cal H}_{\mathrm{int}}$ is invariant under CPT $\equiv \zeta: \; \zeta {{\cal H}_{\mathrm{int}} }\zeta^{-1}= {\cal H}_{\mathrm{int}}$. Taking into account that CPT is an antiunitary operator, 
\begin{eqnarray}
 | \langle \nu_4(\theta, \lambda_4) \,\nu_{\rm light} (\pi-\theta,\lambda_\nu) \;|\; {{{\cal H}_{\mathrm{int}}}}\; |\; Z(\mathrm{\Uparrow}) \rangle | ^2   &=&
  |\;\langle \zeta {{\cal H}_{\mathrm{int}} } \zeta^{-1}\zeta Z(\mathrm{\Uparrow}) \;|\; \zeta \nu_4(\theta, \lambda_4) \, \nu_{\rm light} (\pi-\theta,\lambda_\nu)\rangle \;|^2   \nonumber \\
   &=& |\; \langle {{\cal H}_{\mathrm{int}} } Z(\mathrm{\Downarrow}) \;|\; \nu_4(\theta, -\lambda_4) \, \nu_{\rm light}(\pi-\theta,-\lambda_\nu)\rangle \;|^2   \nonumber \\
  &=& |\;  \langle \nu_4(\pi-\theta, -\lambda_4) \,\nu_{\rm light}(\theta,-\lambda_\nu) \;|\;  {\cal H}_{\mathrm{int}}\; |\;  Z(\mathrm{\Uparrow}) \rangle \;|^2 \;\; ,
\label{eq:CPT}
\end{eqnarray}
where $\Downarrow$ indicates an eigenstate of $S_Z$ with eigenvalue $-1$ and we took advantage of the fact that all three particles are self-conjugate: up to a phase, $Z=\bar{Z}$, $\nu_4=\bar{\nu}_4$, $\nu_{\rm light}=\bar{\nu}_{\rm light}$. In the last step, we assume invariance under a 180$^\circ$-rotation about the axis perpendicular to the decay plane. Summing over the final-state helicities $\lambda_4,\lambda_\nu$, the probability that $\nu_4$ is emitted in the direction $\theta$ is equal to the probability it is emitted in the direction $\pi-\theta$: the decay is forward-backward symmetric. 

\begin{figure}[ht]
\includegraphics[width=0.55\textwidth]{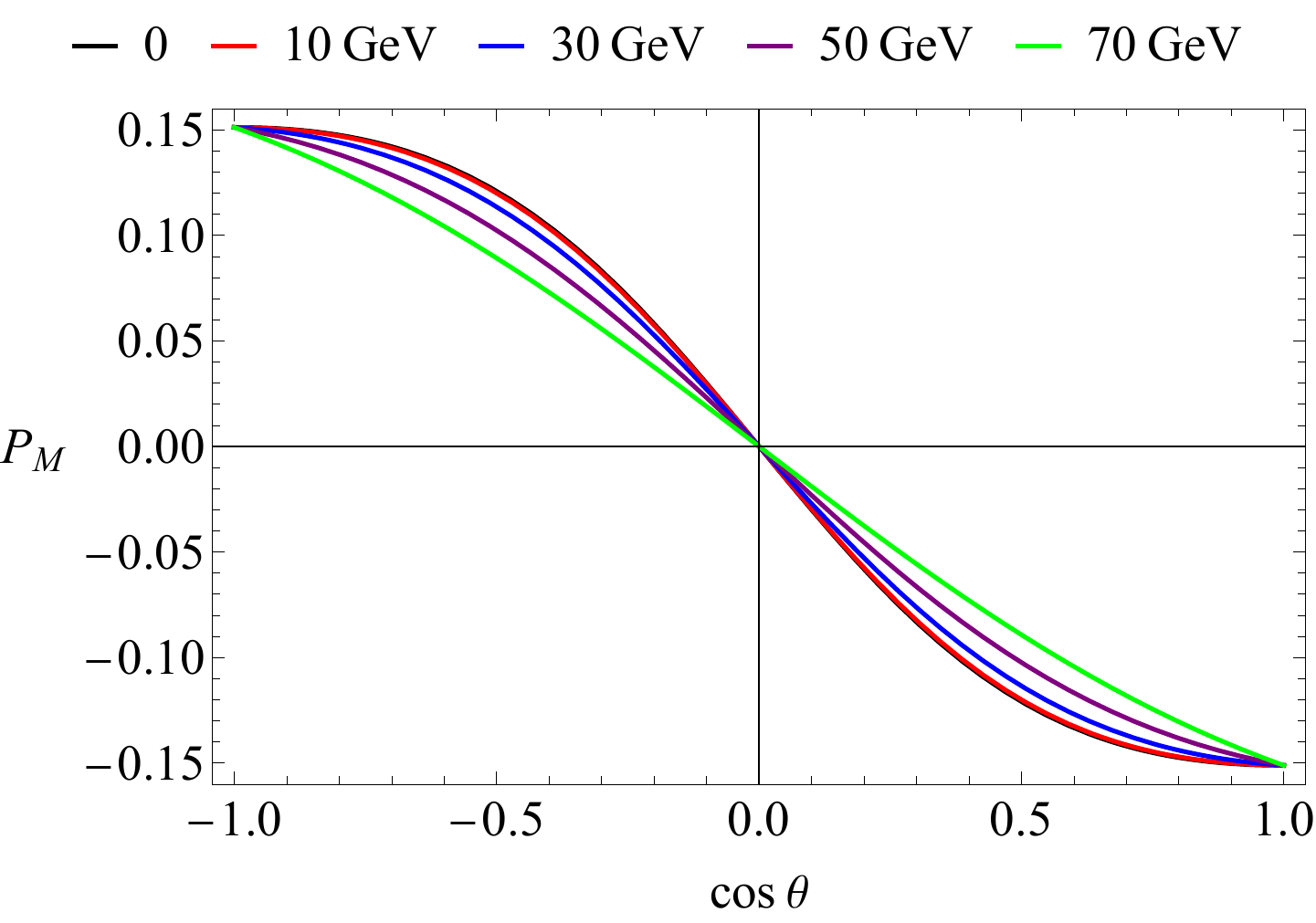}
\caption{The polarizarion  $P_M$ of heavy neutrinos produced in $e^+e^-\to Z \to \nu_4\nu_{\rm light}$ as a function of the direction of the heavy neutrino $\cos\theta$, for different values of the heavy neutrino mass $m_4$. The neutrinos are assumed to be Majorana fermions. The range of $P_M$ values here is much smaller than that of $P_D$ values in Fig.~\ref{fig:PD}.} 
\label{fig:PM}
\end{figure}

While, for any value of $m_4$, the production of heavy Majorana neutrinos with both helicities is allowed by the left-chiral nature of the weak interactions, the probability of emitting a particular helicity eigenstate can depend on the production angle $\theta$. The reason for this is that the $Z$-boson itself is polarized. We find 
\begin{equation}
P_M(\cos\theta)=  \frac{(g_R^2-g_L^2)}{(g_L^2+g_R^2)}\frac{2\cos\theta}{\left(1+\cos^2\theta +\frac{m_4^2}{M_Z^2}\sin^2\theta
\right)}~,
\end{equation}
where the $M$ subscript is to indicate this applies to Majorana neutrinos. $P_M(\cos\theta)$ is depicted in Fig.~\ref{fig:PM}. It is clearly  proportional to the $Z$-boson polarization, Eq.~(\ref{eq:PZ}) and much smaller, in magnitude, than the polarization of Dirac heavy (anti)neutrinos, Fig.~\ref{fig:PD}.

To summarize, in $Z$-decays, Majorana- and Dirac-neutrino production are different. Majorana- and Dirac-neutrino angular distributions (compare Fig.~\ref{fig:DistD} and Fig.~\ref{fig:DistM}) are different and so are their polarization states (compare Fig.~\ref{fig:PD} and Fig.~\ref{fig:PM}). This implies it may be possible, experimentally, to determine the nature of $\nu_4$ produced in $Z$-decay. Of course, the fact that these distributions are different is not sufficient since the $\nu_4$ are not directly observed in the laboratory. Here we assume the $\nu_4$ decay inside the detector and assume their decay products are charged particles which can be identified in a collider experiment. We investigate whether there is enough information to distinguish the hypotheses that neutrinos are Majorana or Dirac fermions in the next section.

\section{Heavy Neutrino Decays and Distinguishing Majorana from Dirac Neutrinos}
\label{sec:Ndecay}
\setcounter{equation}{0}

Heavy neutrino decay is mediated by the charged-current and neutral-current weak interactions. For neutrinos with masses below that of the $W$-boson, one can state that $\nu_4$ decays into either $\ell_{\alpha}+W^*$ or $\nu_i+Z^*$. The neutral-current decay mode is more rare and contains neutrinos in the final state. These are bound to manifest themselves as missing energy, rendering the reconstruction of the $\nu_i\nu_4$ state from $Z$-decay more challenging; the neutrinos also prevent one from determining the total lepton-number of the $\nu_4$ decay products. The charged-current decay mode is more common and less challenging. Except for the cases where the $W^*$ manifests itself as leptons, the entire final state is potentially visible and one can fully reconstruct the four-momentum of the decaying heavy neutrino. In the charged-current decay, the total lepton-number of the $\nu_4$ decay products can also be measured. Henceforth, we discuss exclusively the charged-current decay mode, unless otherwise noted.

If the heavy neutrinos are Dirac fermions, lepton number is conserved: $\nu_4\to \ell^-_{\alpha}(W^*)^+$ while  $\bar{\nu}_4\to \ell^+_{\alpha}(W^*)^-$. If the neutrinos are Majorana fermions, both decay modes are accessible with the same partial width (at leading order): both $\nu_4\to \ell^-_{\alpha}(W^*)^+$ and  $\nu_4\to \ell^+_{\alpha}(W^*)^-$ occur. This fact, by itself, does not allow to distinguish Majorana from Dirac neutrinos produced in $Z$-decay. If the neutrinos are Dirac fermions, the following two decay-chains occur with the same probability:
\begin{eqnarray}
e^+e^-\to Z \to \nu_4 \bar{\nu}_i \to \ell^-(W^*)^+\bar{\nu}_i~, \\
e^+e^-\to Z \to \bar{\nu}_4 \nu_i \to \ell^+(W^*)^- \nu_i~.
\end{eqnarray}
Instead, if the neutrinos are Majorana fermions, the following two decay-chains occur with the same probability:
\begin{eqnarray}
e^+e^-\to Z \to \nu_4 \nu_i \to \ell^-(W^*)^+ \nu_i~, \\
e^+e^-\to Z \to \nu_4 \nu_i \to \ell^+(W^*)^- \nu_i~.
\end{eqnarray}
At the end of the day, if one ``counts'' positively and negatively charged leptons and the light neutrinos escape undetected -- as is (virtually) always the case -- the two scenarios look, globally, identical. 

In more detail, however, the situation is more promising. In the Dirac fermion case, the $\ell^-$ come from parents that are produced with a nonzero $A_{\rm FB}$, Eq.~(\ref{eq:AFBD}), while the $\ell^+$ come from parents that are produced with the opposite nonzero $A_{\rm FB}$, Eq.~(\ref{eq:antiAFBD}). In the Majorana fermion case, both the $\ell^+$ and the $\ell^-$, independently, come from parents that are produced with zero $A_{\rm FB}$, as demonstrated in Sec.~\ref{sec:Zdecay}. Therefore, given a large enough daughter-$\ell^+$ and daughter-$\ell^-$ final-state sample one can distinguish the nature of the neutrinos. They are Dirac fermions if the daughter-$\ell^+$ and daughter-$\ell^-$ distributions reflect nonzero and opposite $A_{FB}$, they are Majorana fermions if both distributions reflect a zero forward-backward asymmetry. We note that the production of heavy neutrinos with masses above 100~GeV at high-energy electron--positron colliders, and whether their nature can be determined using kinematic variables, has been explored in \cite{delAguila:2005pin,Hernandez:2018cgc}. Similar to what we report here, there are kinematic production asymmetries one can exploit when the heavy neutrinos are produced in processes involving off-shell $W$- and $Z$-bosons.

\begin{figure}
\includegraphics[width=0.55\textwidth]{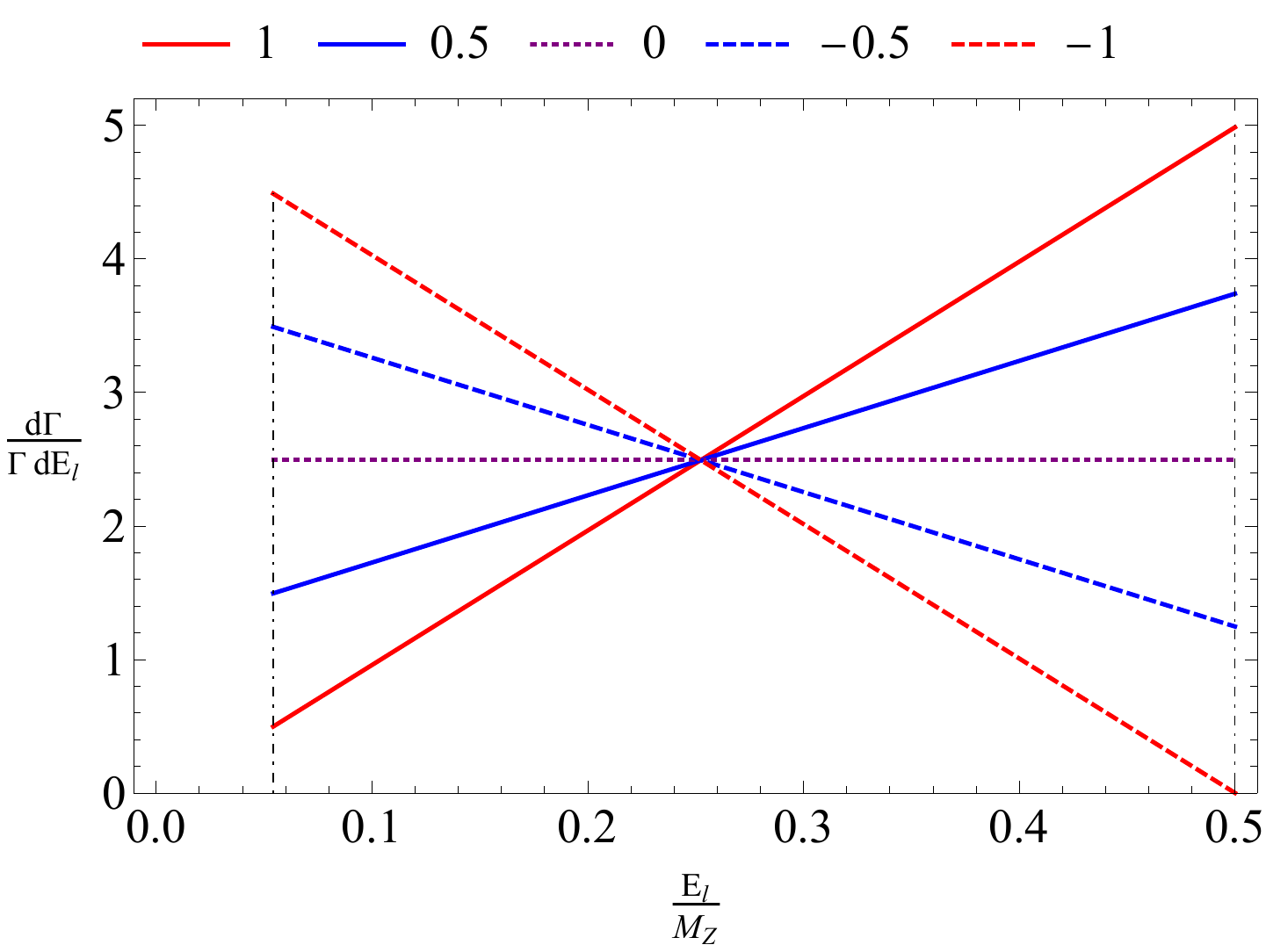}
\caption{Normalized differential decay widths of $\nu_4\to \ell^{\pm}\pi^{\mp}$ as a function of the energy of the charged-lepton, for $\nu_4$ produced in $Z$-decay-at-rest. The different curves correspond to different values of $\alpha_{\pm}P\in [-1,1]$ and $m_4=30$~GeV. Se Eq.~(\ref{eq:Ndecay}).}
\label{fig:Ndecays}
\end{figure}

Another important distinction is that the parent $\nu_4$ and $\bar{\nu}_4$ are both strongly polarized if the neutrinos are Dirac fermions (Fig.~\ref{fig:PD}). Instead, the Majorana-$\nu_4$ are only modestly polarized  (Fig.~\ref{fig:PM}). The kinematics of the final states of the heavy-neutrino decay depend strongly on the parent polarization. As a concrete, simple example, we consider the two-body decay $\nu_4\to \ell_{\alpha}\pi$. In the limit where both the charged-lepton and pion masses are negligible -- a good approximation if $m_4\gtrsim 10$~GeV --  for $\nu_4$ produced in $Z$-boson decay at rest,
\begin{eqnarray}
\frac{1}{\Gamma(\ell^{\pm})}\frac{{\rm d}\Gamma(\ell^{\pm})}{{\rm d}E_{\ell}} = \frac{4}{\left(1-\frac{m_4^2}{M_Z^2}\right)^2}\left[\frac{(1-\alpha_{\pm}P)}{2}-\frac{m_4^2}{M_Z^2}\frac{(1+\alpha_{\pm}P)}{2}+2\alpha_{\pm}P\frac{E_{\ell}}{M_Z}\right]~, \label{eq:Ndecay}
\end{eqnarray}
where $\Gamma(\ell^+)$ and $\Gamma(\ell^-)$ refer, respectively, to the $\ell^+\pi^-$ and $\ell^-\pi^+$ final states and $E_{\ell}\in [m_4^2/(2M_Z),M_Z/2]$ is the charged-lepton energy in the $Z$-boson rest frame. $\alpha_{\pm}$ are the decay-asymmetry parameters, a measure of parity violation in the weak-interactions. Here, $\alpha_+=-\alpha_-=1$. Eq.~(\ref{eq:Ndecay}) applies to both the Majorana-  and Dirac-neutrino cases. In the Dirac-neutrino case, one decay mode is accessible to the heavy antineutrino, the other to the heavy neutrino. We depict these decay distributions in Fig.~\ref{fig:Ndecays} for different values of the heavy neutrino polarization $P$ and $m_4=30$~GeV.

\begin{figure}[ht]
\includegraphics[width=0.49\textwidth]{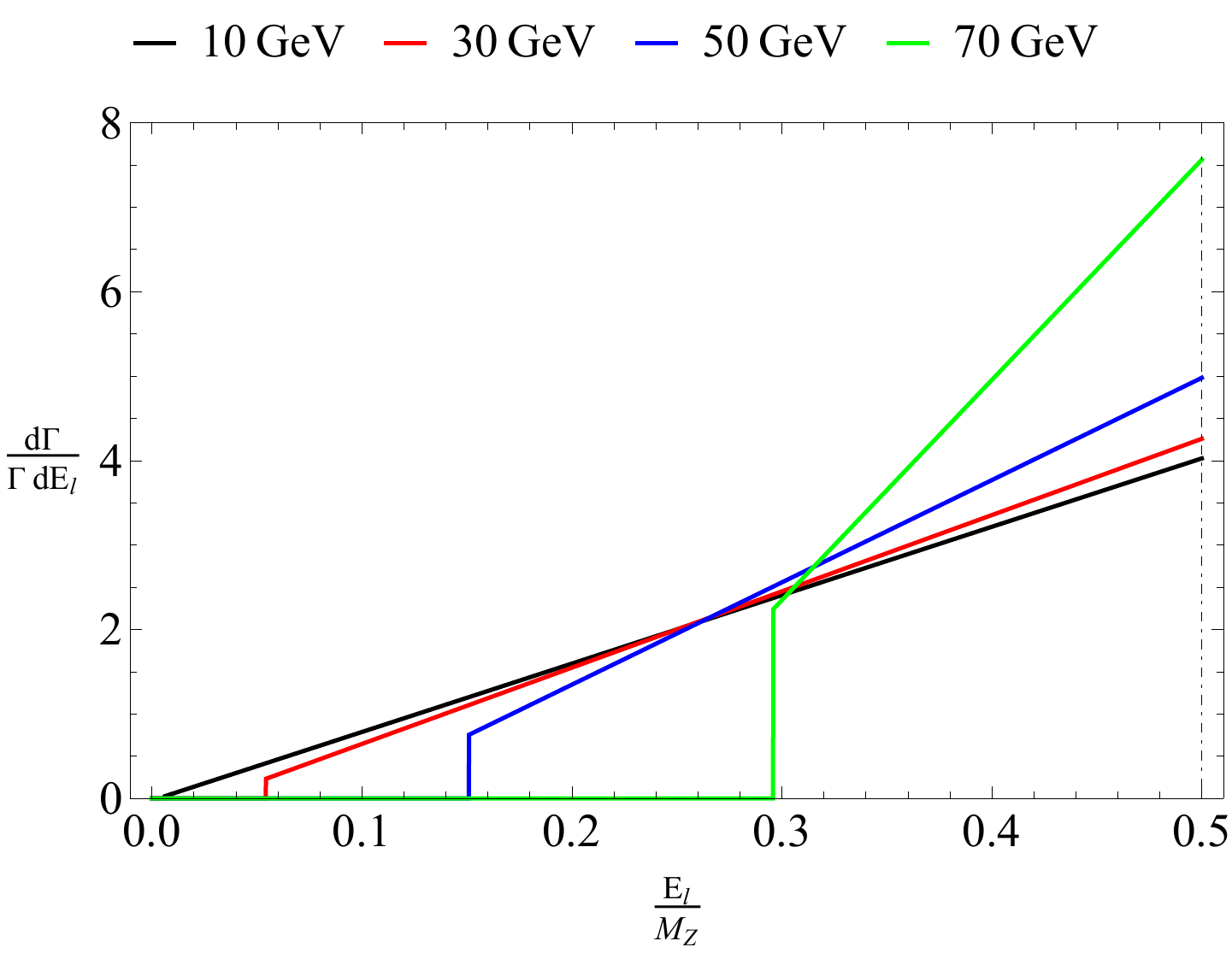}
\includegraphics[width=0.49\textwidth]{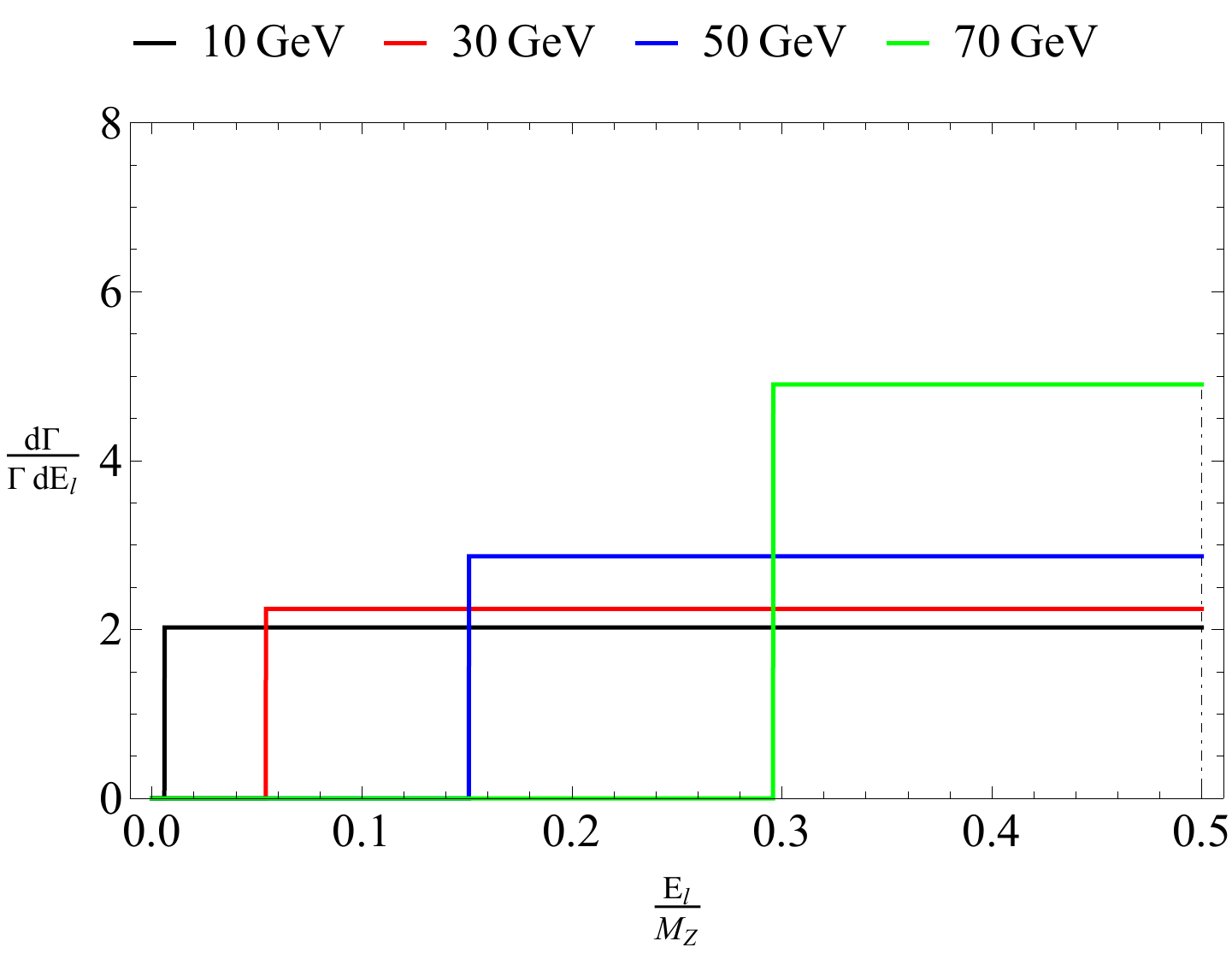}
\caption{Averaged, normalized differential decay widths of  $\nu_4\to\ell^{-}\pi^{+}$ as a function of the energy of the charged-lepton, averaged over the heavy-neutrino production angle, for $\nu_4$ produced in $Z$-decay-at-rest assuming the heavy neutrinos are Dirac (left) and Majorana (right) fermions. The different curves correspond to different values of $m_4$. The same curves apply, both in the left-hand and in the right-hand panels, to the $\ell^{+}\pi^{-}$ final-states.} 
\label{fig:Ndecays_integral}
\end{figure}

The Majorana-fermion or Dirac-fermion nature of the heavy neutrino leaves, therefore, an imprint in the energy distribution of the final-state leptons. Fig.~\ref{fig:Ndecays_integral} depicts the $\ell^+$ and $\ell^-$ energy distributions from $\nu_4$ (and $\bar{\nu}_4$, in the Dirac fermion case) integrated over all $\nu_4$ production angles $\theta$, for both hypotheses concerning the nature of the neutrinos and for different values of $m_4$. In the Dirac-fermion case, the spectra of both $\ell^+$ and $\ell^-$ are ``hard'' (Fig.~\ref{fig:Ndecays_integral}(left)) while in the Majorana-fermion case they are flat (Fig.~\ref{fig:Ndecays_integral}(right)). We highlight here that in order to distinguish Majorana neutrinos from Dirac neutrinos using the charged-lepton energy spectrum, one need-not reconstruct the charge of the charged leptons.  

The $\ell_{\alpha}\pi$ final state is, for $m_4$ values above several GeV, relatively rare, well below a percent for $m_4>10$~GeV, see, for example, \cite{Dib:2018iyr}.\footnote{Very naively we estimate, $B(\nu_4\to\ell_{\alpha}\pi)\sim 5\%(m_{\tau}/m_4)^2$ for $m_4$ masses larger than several GeV and lighter than the $W$-boson mass.} The polarization-dependency of the daughter-charged-lepton energy spectrum is not, however, special to the $\ell_{\alpha}\pi$ two-body final state. There are several other two-body decay modes including two-body decays into other (pseudo)scalar mesons, $\ell_{\alpha}\rho$, and two-body decays into other vector mesons. The branching ratio into $\ell_{\alpha}\rho$ is expected to be much larger than the one into (pesudo)scalar mesons. In the limit $m_4\gg m_{\rho}$, $\alpha_{\pm}=\pm 1$ \cite{Balantekin:2018ukw} so the results discussed above apply directly. 

For relatively heavy $m_4\gtrsim 10$~GeV, well above the narrow hadronic resonances, one can naively consider the $\nu_4$ to decay into $\ell_{\alpha}$ plus ``jets,'' $\nu_4\to\ell_{\alpha}q\bar{q}$. For $\nu_4$ light enough that one can safely integrate out the $W$-boson, one can ``map'' the kinematics of the three-body $\nu_4\to\ell_{\alpha}q\bar{q}$ decay into the well-known Michel decay of the muon. In the $\nu_4$ decay, the charged-lepton plays the role of the muon-neutrino in the Michel decay. Hence the differential decay distribution of the charged lepton from $\nu_4\to\ell_{\alpha}q\bar{q}$ is peaked towards the maximum energy (here, $M_Z/2$) for $\ell^+$ and positive polarization or $\ell^-$ and negative polarization. Instead, it vanishes at the maximum energy for $\ell^+$ and negative polarization or $\ell^-$ and positive polarization, peaking at some intermediate energy. For $P=0$, it falls somewhere between these two limiting cases as depicted, for example, in Figure~9 of~\cite{Albright:2004iw}. A detailed study of all accessible, visible $\nu_4$ decays is beyond the ambitions of this manuscript. Detailed discussions can be found in, for example, \cite{Gorbunov:2007ak,Asaka:2012bb,Ballett:2016opr,Berryman:2017twh,Coloma:2017ppo,Bondarenko:2018ptm,Curtin:2018mvb,SHiP:2018xqw,Ariga:2018uku,Krasnov:2019kdc,Abe:2019kgx,Ballett:2019bgd,Drewes:2019byd,Arguelles:2019ziu,Abratenko:2019kez,Berryman:2019dme,Gorbunov:2020rjx,Coloma:2020lgy,Batell:2020vqn,deVries:2020qns,Plestid:2020ssy,Breitbach:2021gvv}.

It is interesting to appreciate the role of the heavy neutrino decay in the analysis and in our ability to distinguish the Majorana-fermion hypothesis from the Dirac-fermion hypothesis. Imagine, for example, that the heavy neutrinos were very light and very long-lived but one constructed a very efficient, humongous neutrino detector all around the collider experiment. In this case, assuming $|U_{\alpha 4}|^2$ were tiny, it would be best to measure the scattering of the light neutrinos. The distributions computed here apply to the light-neutrinos as well: if they are Dirac-fermions, they are produced with nonzero $A_{FB}$ while if they are Majorana-fermions $A_{FB}$ vanishes, and if they are Dirac-fermions, both the neutrinos and antineutrinos are 100\% polarized with equal and opposite helicity while if they are Majorana-fermions, their polarization is at most as large a that of the $Z$-boson (see Fig.~\ref{fig:PM}). While all these differences are present, one would not be able to distinguish the Majorana-fermion from the Dirac-fermion hypothesis by measuring the neutrinos via scattering. The reason is that these neutrinos are ultra-relativistic. When the left-helicity Majorana-neutrino scatters via charged-current interactions, it behaves exactly like a Dirac-neutrino, i.e., it only produces negatively-charged leptons, and when the right-helicity Majorana-neutrino scatters via charged-current interactions, it behaves exactly like a Dirac-antineutrino, i.e., it only produces positively-charged leptons. The angular distribution of right-handed Majorana-neutrinos is the same as that of Dirac-antineutrinos and that of left-handed Majorana-neutrinos is the same as that of Dirac-neutrinos so the result of the scattering experiment is the same regardless of the nature of the neutrinos.

\section{Concluding Remarks}
\label{sec:conclusion}
\setcounter{equation}{0}

Information on the nature of the neutrinos -- Dirac fermion versus Majorana fermion -- are among the most valuable pieces for solving the neutrino mass puzzle. The same goes for the discovery of new particles. New neutrino-like states are often associated with the new physics responsible for nonzero neutrino masses. Depending on their masses and couplings, these may be produced and detected in a variety of experimental setups. New neutral fermions are not, of course, necessarily directly related to the origin of neutrino masses. Once one of these is discovered, measurements of their properties, including their nature, will be required in order to establish whether they have anything to do with the new physics uncovered by neutrino oscillation experiments. 

We computed the kinematics of $Z$-boson decay into a heavy--light neutrino pair when the $Z$-boson is produced at rest in $e^+e^-$ collisions, including the subsequent decay of the heavy neutrino into a visible final state containing charged-leptons.  We were interested in heavy neutrinos -- defined in Sec.~\ref{sec:heavynu} -- with masses $m_4\sim[10~{\rm GeV},80~{\rm GeV}]$ that decay fast enough that their decay products can be detected in a future Tera-$Z$-like experimental setup. We were especially interested in addressing the nature of the neutrinos, heavy and light, using this physics process.

Majorana- and Dirac-neutrino production are very different. Majorana- and Dirac-neutrino angular distributions (compare Fig.~\ref{fig:DistD} and Fig.~\ref{fig:DistM}) are different and so are their polarization states (compare Fig.~\ref{fig:PD} and Fig.~\ref{fig:PM}). Majorana- and Dirac-neutrino decays are also very different. Heavy Dirac-neutrinos decay into final states with lepton number $+1$, like $\mu^-\pi^+$ while heavy Dirac-antineutrinos decay into final states with lepton number $-1$, like $e^+\pi^-$. Instead, Majorana neutrinos decay into final states with lepton number $+1$ or $-1$ with equal probability (at the leading order); for example, both $\mu^-\rho^+$ and $\mu^+\rho^-$ final states are allowed and equiprobable. Hence, if the heavy neutrinos are Dirac fermions, negatively-charged leptons from heavy-neutrino decay are produced with a nonzero forward-backward asymmetry $A_{\rm FB}$, Eq.~(\ref{eq:AFBD}), while positively-charged  leptons are produced with the opposite nonzero $A_{\rm FB}$, Eq.~(\ref{eq:antiAFBD}). If the heavy neutrinos are Majorana fermions, both the $\ell^+$ and the $\ell^-$ from heavy neutrino decay, independently, are produced with zero $A_{\rm FB}$. One can distinguish the nature of the neutrinos in this way: they are Dirac-fermions if the daughter-$\ell^+$ and daughter-$\ell^-$ distributions reflect nonzero and opposite $A_{FB}$, and they are Majorana-fermions if both distributions reflect a zero forward-backward asymmetry. Moreover, the decay spectrum of the daughter charged-leptons depends on the polarization of the parent which, in turn, depends on the nature of the parent. For the specific case of two-body decays into charged-leptons, we showed that the energy spectra of the daughter charged-leptons are very different even if we integrate over the kinematics of the parent heavy-neutrino, as depicted in Fig.~\ref{fig:Ndecays_integral}. 

In summary, given enough events, it is possible to distinguish whether the hypothetical heavy--light neutrino pairs produced in a $Z$-factory are Majorana fermions or Dirac fermions, even though the violation of lepton-number, or lack thereof, cannot be established in an event-by-event basis. A quantitative appraisal of the sensitivity at proposed next-generation lepton colliders, taking into account, for example, different decay modes and the finite lifetime of the heavy neutrino, is left for future work.

\section*{Acknowledgements}
We thank Kevin Kelly for many useful comments on the manuscript and Matthew Mccullough and Gaia Lanfranchi for stimulating discussions. This work was supported in part by the US Department of Energy (DOE) grant \#de-sc0010143 and in part by the NSF grant PHY-1630782. The document was prepared using the resources of the Fermi National Accelerator Laboratory (Fermilab), a DOE, Office of Science, HEP User Facility. Fermilab is managed by Fermi Research Alliance, LLC (FRA), acting under Contract No. DE-AC02-07CH11359.

\bibliographystyle{apsrev-title}
\bibliography{ZtoNnu.bib}{}

\begin{thebibliography}{65}
\expandafter\ifx\csname natexlab\endcsname\relax\def\natexlab#1{#1}\fi
\expandafter\ifx\csname bibnamefont\endcsname\relax
  \def\bibnamefont#1{#1}\fi
\expandafter\ifx\csname bibfnamefont\endcsname\relax
  \def\bibfnamefont#1{#1}\fi
\expandafter\ifx\csname citenamefont\endcsname\relax
  \def\citenamefont#1{#1}\fi
\expandafter\ifx\csname url\endcsname\relax
  \def\url#1{\texttt{#1}}\fi
\expandafter\ifx\csname urlprefix\endcsname\relax\def\urlprefix{URL }\fi
\providecommand{\bibinfo}[2]{#2}
\providecommand{\eprint}[2][]{\url{#2}}

\bibitem[{\citenamefont{de~Gouv\^ea and Vogel}(2013)}]{deGouvea:2013zba}
\bibinfo{author}{\bibfnamefont{A.}~\bibnamefont{de~Gouv\^ea}} \bibnamefont{and}
  \bibinfo{author}{\bibfnamefont{P.}~\bibnamefont{Vogel}}, ``{Lepton Flavor and
  Number Conservation, and Physics Beyond the Standard Model},''
  \bibinfo{journal}{Prog. Part. Nucl. Phys.} \textbf{\bibinfo{volume}{71}},
  \bibinfo{pages}{75} (\bibinfo{year}{2013}), \eprint{1303.4097}.

\bibitem[{\citenamefont{Dolinski et~al.}(2019)\citenamefont{Dolinski, Poon, and
  Rodejohann}}]{Dolinski:2019nrj}
\bibinfo{author}{\bibfnamefont{M.~J.} \bibnamefont{Dolinski}},
  \bibinfo{author}{\bibfnamefont{A.~W.} \bibnamefont{Poon}}, \bibnamefont{and}
  \bibinfo{author}{\bibfnamefont{W.}~\bibnamefont{Rodejohann}}, ``{Neutrinoless
  Double-Beta Decay: Status and Prospects},'' \bibinfo{journal}{Ann. Rev. Nucl.
  Part. Sci.} \textbf{\bibinfo{volume}{69}}, \bibinfo{pages}{219}
  (\bibinfo{year}{2019}), \eprint{1902.04097}.

\bibitem[{\citenamefont{Schechter and Valle}(1982)}]{Schechter:1981bd}
\bibinfo{author}{\bibfnamefont{J.}~\bibnamefont{Schechter}} \bibnamefont{and}
  \bibinfo{author}{\bibfnamefont{J.~W.~F.} \bibnamefont{Valle}},
  ``{Neutrinoless Double beta Decay in SU(2) x U(1) Theories},''
  \bibinfo{journal}{Phys. Rev. D} \textbf{\bibinfo{volume}{25}},
  \bibinfo{pages}{2951} (\bibinfo{year}{1982}).

\bibitem[{\citenamefont{Nieves}(1984)}]{Nieves:1984sn}
\bibinfo{author}{\bibfnamefont{J.~F.} \bibnamefont{Nieves}}, ``{Dirac and
  Pseudodirac Neutrinos and Neutrinoless Double Beta Decay},''
  \bibinfo{journal}{Phys. Lett. B} \textbf{\bibinfo{volume}{147}},
  \bibinfo{pages}{375} (\bibinfo{year}{1984}).

\bibitem[{\citenamefont{Takasugi}(1984)}]{Takasugi:1984xr}
\bibinfo{author}{\bibfnamefont{E.}~\bibnamefont{Takasugi}}, ``{Can the
  Neutrinoless Double Beta Decay Take Place in the Case of Dirac Neutrinos?},''
  \bibinfo{journal}{Phys. Lett. B} \textbf{\bibinfo{volume}{149}},
  \bibinfo{pages}{372} (\bibinfo{year}{1984}).

\bibitem[{\citenamefont{Hirsch et~al.}(2006)\citenamefont{Hirsch, Kovalenko,
  and Schmidt}}]{Hirsch:2006yk}
\bibinfo{author}{\bibfnamefont{M.}~\bibnamefont{Hirsch}},
  \bibinfo{author}{\bibfnamefont{S.}~\bibnamefont{Kovalenko}},
  \bibnamefont{and} \bibinfo{author}{\bibfnamefont{I.}~\bibnamefont{Schmidt}},
  ``{Extended black box theorem for lepton number and flavor violating
  processes},'' \bibinfo{journal}{Phys. Lett. B}
  \textbf{\bibinfo{volume}{642}}, \bibinfo{pages}{106} (\bibinfo{year}{2006}),
  \eprint{hep-ph/0608207}.

\bibitem[{\citenamefont{Zyla et~al.}(2020)}]{Zyla:2020zbs}
\bibinfo{author}{\bibfnamefont{P.~A.} \bibnamefont{Zyla}} \bibnamefont{et~al.}
  (\bibinfo{collaboration}{Particle Data Group}), ``{Review of Particle
  Physics},'' \bibinfo{journal}{PTEP} \textbf{\bibinfo{volume}{2020}},
  \bibinfo{pages}{083C01} (\bibinfo{year}{2020}).

\bibitem[{\citenamefont{de~Gouv\^ea}(2016)}]{Gouvea:2016shl}
\bibinfo{author}{\bibfnamefont{A.}~\bibnamefont{de~Gouv\^ea}}, ``{Neutrino Mass
  Models},'' \bibinfo{journal}{Ann. Rev. Nucl. Part. Sci.}
  \textbf{\bibinfo{volume}{66}}, \bibinfo{pages}{197} (\bibinfo{year}{2016}).

\bibitem[{\citenamefont{Minkowski}(1977)}]{Minkowski:1977sc}
\bibinfo{author}{\bibfnamefont{P.}~\bibnamefont{Minkowski}}, ``{$\mu \to
  e\gamma$ at a Rate of One Out of $10^{9}$ Muon Decays?},''
  \bibinfo{journal}{Phys. Lett. B} \textbf{\bibinfo{volume}{67}},
  \bibinfo{pages}{421} (\bibinfo{year}{1977}).

\bibitem[{\citenamefont{Gell-Mann et~al.}(1979)\citenamefont{Gell-Mann, Ramond,
  and Slansky}}]{GellMann:1980vs}
\bibinfo{author}{\bibfnamefont{M.}~\bibnamefont{Gell-Mann}},
  \bibinfo{author}{\bibfnamefont{P.}~\bibnamefont{Ramond}}, \bibnamefont{and}
  \bibinfo{author}{\bibfnamefont{R.}~\bibnamefont{Slansky}}, ``{Complex Spinors
  and Unified Theories},'' \bibinfo{journal}{Conf. Proc. C}
  \textbf{\bibinfo{volume}{790927}}, \bibinfo{pages}{315}
  (\bibinfo{year}{1979}), \eprint{1306.4669}.

\bibitem[{\citenamefont{Yanagida}(1979)}]{Yanagida:1979as}
\bibinfo{author}{\bibfnamefont{T.}~\bibnamefont{Yanagida}}, ``{Horizontal gauge
  symmetry and masses of neutrinos},'' \bibinfo{journal}{Conf. Proc. C}
  \textbf{\bibinfo{volume}{7902131}}, \bibinfo{pages}{95}
  (\bibinfo{year}{1979}).

\bibitem[{\citenamefont{Glashow}(1980)}]{Glashow:1979nm}
\bibinfo{author}{\bibfnamefont{S.~L.} \bibnamefont{Glashow}}, ``{The Future of
  Elementary Particle Physics},'' \bibinfo{journal}{NATO Sci. Ser. B}
  \textbf{\bibinfo{volume}{61}}, \bibinfo{pages}{687} (\bibinfo{year}{1980}).

\bibitem[{\citenamefont{Mohapatra and Senjanovic}(1980)}]{Mohapatra:1979ia}
\bibinfo{author}{\bibfnamefont{R.~N.} \bibnamefont{Mohapatra}}
  \bibnamefont{and}
  \bibinfo{author}{\bibfnamefont{G.}~\bibnamefont{Senjanovic}}, ``{Neutrino
  Mass and Spontaneous Parity Nonconservation},'' \bibinfo{journal}{Phys. Rev.
  Lett.} \textbf{\bibinfo{volume}{44}}, \bibinfo{pages}{912}
  (\bibinfo{year}{1980}).

\bibitem[{\citenamefont{Schechter and Valle}(1980)}]{Schechter:1980gr}
\bibinfo{author}{\bibfnamefont{J.}~\bibnamefont{Schechter}} \bibnamefont{and}
  \bibinfo{author}{\bibfnamefont{J.~W.~F.} \bibnamefont{Valle}}, ``{Neutrino
  Masses in SU(2) x U(1) Theories},'' \bibinfo{journal}{Phys. Rev. D}
  \textbf{\bibinfo{volume}{22}}, \bibinfo{pages}{2227} (\bibinfo{year}{1980}).

\bibitem[{\citenamefont{de~Gouv\^ea and Hern\'andez}(2015)}]{deGouvea:2015pea}
\bibinfo{author}{\bibfnamefont{A.}~\bibnamefont{de~Gouv\^ea}} \bibnamefont{and}
  \bibinfo{author}{\bibfnamefont{D.}~\bibnamefont{Hern\'andez}}, ``{New Chiral
  Fermions, a New Gauge Interaction, Dirac Neutrinos, and Dark Matter},''
  \bibinfo{journal}{JHEP} \textbf{\bibinfo{volume}{10}}, \bibinfo{pages}{046}
  (\bibinfo{year}{2015}), \eprint{1507.00916}.

\bibitem[{\citenamefont{Arkani-Hamed et~al.}(2001)\citenamefont{Arkani-Hamed,
  Dimopoulos, Dvali, and March-Russell}}]{ArkaniHamed:1998vp}
\bibinfo{author}{\bibfnamefont{N.}~\bibnamefont{Arkani-Hamed}},
  \bibinfo{author}{\bibfnamefont{S.}~\bibnamefont{Dimopoulos}},
  \bibinfo{author}{\bibfnamefont{G.~R.} \bibnamefont{Dvali}}, \bibnamefont{and}
  \bibinfo{author}{\bibfnamefont{J.}~\bibnamefont{March-Russell}}, ``{Neutrino
  masses from large extra dimensions},'' \bibinfo{journal}{Phys. Rev. D}
  \textbf{\bibinfo{volume}{65}}, \bibinfo{pages}{024032}
  (\bibinfo{year}{2001}), \eprint{hep-ph/9811448}.

\bibitem[{\citenamefont{Dienes et~al.}(1999)\citenamefont{Dienes, Dudas, and
  Gherghetta}}]{Dienes:1998sb}
\bibinfo{author}{\bibfnamefont{K.~R.} \bibnamefont{Dienes}},
  \bibinfo{author}{\bibfnamefont{E.}~\bibnamefont{Dudas}}, \bibnamefont{and}
  \bibinfo{author}{\bibfnamefont{T.}~\bibnamefont{Gherghetta}}, ``{Neutrino
  oscillations without neutrino masses or heavy mass scales: A Higher
  dimensional seesaw mechanism},'' \bibinfo{journal}{Nucl. Phys. B}
  \textbf{\bibinfo{volume}{557}}, \bibinfo{pages}{25} (\bibinfo{year}{1999}),
  \eprint{hep-ph/9811428}.

\bibitem[{\citenamefont{Dvali and Smirnov}(1999)}]{Dvali:1999cn}
\bibinfo{author}{\bibfnamefont{G.~R.} \bibnamefont{Dvali}} \bibnamefont{and}
  \bibinfo{author}{\bibfnamefont{A.~Y.} \bibnamefont{Smirnov}}, ``{Probing
  large extra dimensions with neutrinos},'' \bibinfo{journal}{Nucl. Phys. B}
  \textbf{\bibinfo{volume}{563}}, \bibinfo{pages}{63} (\bibinfo{year}{1999}),
  \eprint{hep-ph/9904211}.

\bibitem[{\citenamefont{Balantekin et~al.}(2019)\citenamefont{Balantekin,
  de~Gouv\^ea, and Kayser}}]{Balantekin:2018ukw}
\bibinfo{author}{\bibfnamefont{A.~B.} \bibnamefont{Balantekin}},
  \bibinfo{author}{\bibfnamefont{A.}~\bibnamefont{de~Gouv\^ea}},
  \bibnamefont{and} \bibinfo{author}{\bibfnamefont{B.}~\bibnamefont{Kayser}},
  ``{Addressing the Majorana vs. Dirac Question with Neutrino Decays},''
  \bibinfo{journal}{Phys. Lett. B} \textbf{\bibinfo{volume}{789}},
  \bibinfo{pages}{488} (\bibinfo{year}{2019}), \eprint{1808.10518}.

\bibitem[{\citenamefont{B.~Balantekin and Kayser}(2018)}]{Balantekin:2018azf}
\bibinfo{author}{\bibfnamefont{A.}~\bibnamefont{B.~Balantekin}}
  \bibnamefont{and} \bibinfo{author}{\bibfnamefont{B.}~\bibnamefont{Kayser}},
  ``{On the Properties of Neutrinos},''  (\bibinfo{year}{2018}),
  \eprint{1805.00922}.

\bibitem[{\citenamefont{Shrock}(1982)}]{Shrock:1982sc}
\bibinfo{author}{\bibfnamefont{R.~E.} \bibnamefont{Shrock}}, ``{Electromagnetic
  Properties and Decays of Dirac and Majorana Neutrinos in a General Class of
  Gauge Theories},'' \bibinfo{journal}{Nucl. Phys.}
  \textbf{\bibinfo{volume}{B206}}, \bibinfo{pages}{359} (\bibinfo{year}{1982}).

\bibitem[{\citenamefont{Li and Wilczek}(1982)}]{Li:1981um}
\bibinfo{author}{\bibfnamefont{L.~F.} \bibnamefont{Li}} \bibnamefont{and}
  \bibinfo{author}{\bibfnamefont{F.}~\bibnamefont{Wilczek}}, ``{PHYSICAL
  PROCESSES INVOLVING MAJORANA NEUTRINOS},'' \bibinfo{journal}{Phys. Rev.}
  \textbf{\bibinfo{volume}{D25}}, \bibinfo{pages}{143} (\bibinfo{year}{1982}).

\bibitem[{\citenamefont{Abada et~al.}(2019)}]{Benedikt:2651299}
\bibinfo{author}{\bibfnamefont{A.}~\bibnamefont{Abada}} \bibnamefont{et~al.},
  ``{FCC-ee: The Lepton Collider}; {FCC CDR Vol. 2},'' \bibinfo{journal}{Eur.
  Phys. J. ST} \textbf{\bibinfo{volume}{228}}, \bibinfo{pages}{261}
  (\bibinfo{year}{2019}).

\bibitem[{\citenamefont{Chou et~al.}(2018)}]{CEPCStudyGroup:2018rmc}
\bibinfo{author}{\bibfnamefont{W.}~\bibnamefont{Chou}} \bibnamefont{et~al.}
  (\bibinfo{collaboration}{CEPC Study Group}), ``{CEPC Conceptual Design
  Report: Volume 1 - Accelerator},''  (\bibinfo{year}{2018}),
  \eprint{1809.00285}.

\bibitem[{\citenamefont{\textsc{CEPC Study Group}
  et~al.}(2018)\citenamefont{\textsc{CEPC Study Group}, Dong
  et~al.}}]{CEPCStudyGroup:2018ghi}
\bibinfo{author}{\bibnamefont{\textsc{CEPC Study Group}}},
  \bibinfo{author}{\bibfnamefont{M.}~\bibnamefont{Dong}}, \bibnamefont{et~al.},
  ``{CEPC Conceptual Design Report: Volume 2 - Physics \& Detector},''
  (\bibinfo{year}{2018}), \eprint{1811.10545}.

\bibitem[{\citenamefont{Dittmar et~al.}(1990)\citenamefont{Dittmar, Santamaria,
  Gonzalez-Garcia, and Valle}}]{Dittmar:1989yg}
\bibinfo{author}{\bibfnamefont{M.}~\bibnamefont{Dittmar}},
  \bibinfo{author}{\bibfnamefont{A.}~\bibnamefont{Santamaria}},
  \bibinfo{author}{\bibfnamefont{M.~C.} \bibnamefont{Gonzalez-Garcia}},
  \bibnamefont{and} \bibinfo{author}{\bibfnamefont{J.~W.~F.}
  \bibnamefont{Valle}}, ``{Production Mechanisms and Signatures of Isosinglet
  Neutral Heavy Leptons in $Z^0$ Decays},'' \bibinfo{journal}{Nucl. Phys. B}
  \textbf{\bibinfo{volume}{332}}, \bibinfo{pages}{1} (\bibinfo{year}{1990}).

\bibitem[{\citenamefont{Abreu et~al.}(1997)}]{Abreu:1996pa}
\bibinfo{author}{\bibfnamefont{P.}~\bibnamefont{Abreu}} \bibnamefont{et~al.}
  (\bibinfo{collaboration}{DELPHI}), ``{Search for neutral heavy leptons
  produced in Z decays},'' \bibinfo{journal}{Z. Phys. C}
  \textbf{\bibinfo{volume}{74}}, \bibinfo{pages}{57} (\bibinfo{year}{1997}),
  \bibinfo{note}{[Erratum: Z.Phys.C 75, 580 (1997)]}.

\bibitem[{\citenamefont{Drewes and Garbrecht}(2017)}]{Drewes:2015iva}
\bibinfo{author}{\bibfnamefont{M.}~\bibnamefont{Drewes}} \bibnamefont{and}
  \bibinfo{author}{\bibfnamefont{B.}~\bibnamefont{Garbrecht}}, ``{Combining
  experimental and cosmological constraints on heavy neutrinos},''
  \bibinfo{journal}{Nucl. Phys. B} \textbf{\bibinfo{volume}{921}},
  \bibinfo{pages}{250} (\bibinfo{year}{2017}), \eprint{1502.00477}.

\bibitem[{\citenamefont{Fernandez-Martinez
  et~al.}(2016)\citenamefont{Fernandez-Martinez, Hernandez-Garcia, and
  Lopez-Pavon}}]{Fernandez-Martinez:2016lgt}
\bibinfo{author}{\bibfnamefont{E.}~\bibnamefont{Fernandez-Martinez}},
  \bibinfo{author}{\bibfnamefont{J.}~\bibnamefont{Hernandez-Garcia}},
  \bibnamefont{and}
  \bibinfo{author}{\bibfnamefont{J.}~\bibnamefont{Lopez-Pavon}}, ``{Global
  constraints on heavy neutrino mixing},'' \bibinfo{journal}{JHEP}
  \textbf{\bibinfo{volume}{08}}, \bibinfo{pages}{033} (\bibinfo{year}{2016}),
  \eprint{1605.08774}.

\bibitem[{\citenamefont{Drewes et~al.}(2017)\citenamefont{Drewes, Garbrecht,
  Gueter, and Klaric}}]{Drewes:2016jae}
\bibinfo{author}{\bibfnamefont{M.}~\bibnamefont{Drewes}},
  \bibinfo{author}{\bibfnamefont{B.}~\bibnamefont{Garbrecht}},
  \bibinfo{author}{\bibfnamefont{D.}~\bibnamefont{Gueter}}, \bibnamefont{and}
  \bibinfo{author}{\bibfnamefont{J.}~\bibnamefont{Klaric}}, ``{Testing the low
  scale seesaw and leptogenesis},'' \bibinfo{journal}{JHEP}
  \textbf{\bibinfo{volume}{08}}, \bibinfo{pages}{018} (\bibinfo{year}{2017}),
  \eprint{1609.09069}.

\bibitem[{\citenamefont{Bryman and
  Shrock}(2019{\natexlab{a}})}]{Bryman:2019ssi}
\bibinfo{author}{\bibfnamefont{D.~A.} \bibnamefont{Bryman}} \bibnamefont{and}
  \bibinfo{author}{\bibfnamefont{R.}~\bibnamefont{Shrock}}, ``{Improved
  Constraints on Sterile Neutrinos in the MeV to GeV Mass Range},''
  \bibinfo{journal}{Phys. Rev. D} \textbf{\bibinfo{volume}{100}},
  \bibinfo{pages}{053006} (\bibinfo{year}{2019}{\natexlab{a}}),
  \eprint{1904.06787}.

\bibitem[{\citenamefont{Bryman and
  Shrock}(2019{\natexlab{b}})}]{Bryman:2019bjg}
\bibinfo{author}{\bibfnamefont{D.~A.} \bibnamefont{Bryman}} \bibnamefont{and}
  \bibinfo{author}{\bibfnamefont{R.}~\bibnamefont{Shrock}}, ``{Constraints on
  Sterile Neutrinos in the MeV to GeV Mass Range},'' \bibinfo{journal}{Phys.
  Rev. D} \textbf{\bibinfo{volume}{100}}, \bibinfo{pages}{073011}
  (\bibinfo{year}{2019}{\natexlab{b}}), \eprint{1909.11198}.

\bibitem[{\citenamefont{Bolton et~al.}(2020)\citenamefont{Bolton, Deppisch, and
  Bhupal~Dev}}]{Bolton:2019pcu}
\bibinfo{author}{\bibfnamefont{P.~D.} \bibnamefont{Bolton}},
  \bibinfo{author}{\bibfnamefont{F.~F.} \bibnamefont{Deppisch}},
  \bibnamefont{and}
  \bibinfo{author}{\bibfnamefont{P.}~\bibnamefont{Bhupal~Dev}}, ``{Neutrinoless
  double beta decay versus other probes of heavy sterile neutrinos},''
  \bibinfo{journal}{JHEP} \textbf{\bibinfo{volume}{03}}, \bibinfo{pages}{170}
  (\bibinfo{year}{2020}), \eprint{1912.03058}.

\bibitem[{\citenamefont{Blondel et~al.}(2016)\citenamefont{Blondel, Graverini,
  Serra, and Shaposhnikov}}]{Blondel:2014bra}
\bibinfo{author}{\bibfnamefont{A.}~\bibnamefont{Blondel}},
  \bibinfo{author}{\bibfnamefont{E.}~\bibnamefont{Graverini}},
  \bibinfo{author}{\bibfnamefont{N.}~\bibnamefont{Serra}}, \bibnamefont{and}
  \bibinfo{author}{\bibfnamefont{M.}~\bibnamefont{Shaposhnikov}}
  (\bibinfo{collaboration}{FCC-ee study Team}), ``{Search for Heavy Right
  Handed Neutrinos at the FCC-ee},'' \bibinfo{journal}{Nucl. Part. Phys. Proc.}
  \textbf{\bibinfo{volume}{273-275}}, \bibinfo{pages}{1883}
  (\bibinfo{year}{2016}), \eprint{1411.5230}.

\bibitem[{\citenamefont{Antusch et~al.}(2016)\citenamefont{Antusch, Cazzato,
  and Fischer}}]{Antusch:2016vyf}
\bibinfo{author}{\bibfnamefont{S.}~\bibnamefont{Antusch}},
  \bibinfo{author}{\bibfnamefont{E.}~\bibnamefont{Cazzato}}, \bibnamefont{and}
  \bibinfo{author}{\bibfnamefont{O.}~\bibnamefont{Fischer}}, ``{Displaced
  vertex searches for sterile neutrinos at future lepton colliders},''
  \bibinfo{journal}{JHEP} \textbf{\bibinfo{volume}{12}}, \bibinfo{pages}{007}
  (\bibinfo{year}{2016}), \eprint{1604.02420}.

\bibitem[{\citenamefont{Liao and Wu}(2018)}]{Liao:2017jiz}
\bibinfo{author}{\bibfnamefont{W.}~\bibnamefont{Liao}} \bibnamefont{and}
  \bibinfo{author}{\bibfnamefont{X.-H.} \bibnamefont{Wu}}, ``{Signature of
  heavy sterile neutrinos at CEPC},'' \bibinfo{journal}{Phys. Rev. D}
  \textbf{\bibinfo{volume}{97}}, \bibinfo{pages}{055005}
  (\bibinfo{year}{2018}), \eprint{1710.09266}.

\bibitem[{\citenamefont{Blondel et~al.}(2018)}]{Blondel:2018mad}
\bibinfo{author}{\bibfnamefont{A.}~\bibnamefont{Blondel}} \bibnamefont{et~al.},
  in \emph{\bibinfo{booktitle}{{Mini Workshop on Precision EW and QCD
  Calculations for the FCC Studies : Methods and Techniques}}}
  (\bibinfo{publisher}{CERN}, \bibinfo{address}{Geneva}, \bibinfo{year}{2018}),
  vol. \bibinfo{volume}{3/2019} of \emph{\bibinfo{series}{CERN Yellow Reports:
  Monographs}}, \eprint{1809.01830}.

\bibitem[{\citenamefont{Ding et~al.}(2019)\citenamefont{Ding, Qin, and
  Yu}}]{Ding:2019tqq}
\bibinfo{author}{\bibfnamefont{J.-N.} \bibnamefont{Ding}},
  \bibinfo{author}{\bibfnamefont{Q.}~\bibnamefont{Qin}}, \bibnamefont{and}
  \bibinfo{author}{\bibfnamefont{F.-S.} \bibnamefont{Yu}}, ``{Heavy neutrino
  searches at future $Z$-factories},'' \bibinfo{journal}{Eur. Phys. J. C}
  \textbf{\bibinfo{volume}{79}}, \bibinfo{pages}{766} (\bibinfo{year}{2019}),
  \eprint{1903.02570}.

\bibitem[{\citenamefont{de~Gouv\^ea et~al.}(2021)\citenamefont{de~Gouv\^ea,
  Fox, Kayser, and Kelly}}]{deGouvea:2021ual}
\bibinfo{author}{\bibfnamefont{A.}~\bibnamefont{de~Gouv\^ea}},
  \bibinfo{author}{\bibfnamefont{P.~J.} \bibnamefont{Fox}},
  \bibinfo{author}{\bibfnamefont{B.~J.} \bibnamefont{Kayser}},
  \bibnamefont{and} \bibinfo{author}{\bibfnamefont{K.~J.} \bibnamefont{Kelly}},
  ``{Three-Body Decays of Heavy Dirac and Majorana Fermions},''
  (\bibinfo{year}{2021}), \eprint{2104.05719}.

\bibitem[{\citenamefont{del Aguila and
  Aguilar-Saavedra}(2005)}]{delAguila:2005pin}
\bibinfo{author}{\bibfnamefont{F.}~\bibnamefont{del Aguila}} \bibnamefont{and}
  \bibinfo{author}{\bibfnamefont{J.~A.} \bibnamefont{Aguilar-Saavedra}}, ``{l W
  nu production at CLIC: A Window to TeV scale non-decoupled neutrinos},''
  \bibinfo{journal}{JHEP} \textbf{\bibinfo{volume}{05}}, \bibinfo{pages}{026}
  (\bibinfo{year}{2005}), \eprint{hep-ph/0503026}.

\bibitem[{\citenamefont{Hern\'andez et~al.}(2019)\citenamefont{Hern\'andez,
  Jones-P\'erez, and Suarez-Navarro}}]{Hernandez:2018cgc}
\bibinfo{author}{\bibfnamefont{P.}~\bibnamefont{Hern\'andez}},
  \bibinfo{author}{\bibfnamefont{J.}~\bibnamefont{Jones-P\'erez}},
  \bibnamefont{and}
  \bibinfo{author}{\bibfnamefont{O.}~\bibnamefont{Suarez-Navarro}}, ``{Majorana
  vs Pseudo-Dirac Neutrinos at the ILC},'' \bibinfo{journal}{Eur. Phys. J. C}
  \textbf{\bibinfo{volume}{79}}, \bibinfo{pages}{220} (\bibinfo{year}{2019}),
  \eprint{1810.07210}.

\bibitem[{\citenamefont{Dib et~al.}(2018)\citenamefont{Dib, Kim, Neill, and
  Yuan}}]{Dib:2018iyr}
\bibinfo{author}{\bibfnamefont{C.~O.} \bibnamefont{Dib}},
  \bibinfo{author}{\bibfnamefont{C.~S.} \bibnamefont{Kim}},
  \bibinfo{author}{\bibfnamefont{N.~A.} \bibnamefont{Neill}}, \bibnamefont{and}
  \bibinfo{author}{\bibfnamefont{X.-B.} \bibnamefont{Yuan}}, ``{Search for
  sterile neutrinos decaying into pions at the LHC},'' \bibinfo{journal}{Phys.
  Rev. D} \textbf{\bibinfo{volume}{97}}, \bibinfo{pages}{035022}
  (\bibinfo{year}{2018}), \eprint{1801.03624}.

\bibitem[{\citenamefont{Albright et~al.}(2004)}]{Albright:2004iw}
\bibinfo{author}{\bibfnamefont{C.}~\bibnamefont{Albright}} \bibnamefont{et~al.}
  (\bibinfo{collaboration}{Neutrino Factory, Muon Collider}), ``{The neutrino
  factory and beta beam experiments and development},''
  (\bibinfo{year}{2004}), \eprint{physics/0411123}.

\bibitem[{\citenamefont{Gorbunov and Shaposhnikov}(2007)}]{Gorbunov:2007ak}
\bibinfo{author}{\bibfnamefont{D.}~\bibnamefont{Gorbunov}} \bibnamefont{and}
  \bibinfo{author}{\bibfnamefont{M.}~\bibnamefont{Shaposhnikov}}, ``{How to
  find neutral leptons of the $\nu$MSM?},'' \bibinfo{journal}{JHEP}
  \textbf{\bibinfo{volume}{10}}, \bibinfo{pages}{015} (\bibinfo{year}{2007}),
  \bibinfo{note}{[Erratum: JHEP 11, 101 (2013)]}, \eprint{0705.1729}.

\bibitem[{\citenamefont{Asaka et~al.}(2013)\citenamefont{Asaka, Eijima, and
  Watanabe}}]{Asaka:2012bb}
\bibinfo{author}{\bibfnamefont{T.}~\bibnamefont{Asaka}},
  \bibinfo{author}{\bibfnamefont{S.}~\bibnamefont{Eijima}}, \bibnamefont{and}
  \bibinfo{author}{\bibfnamefont{A.}~\bibnamefont{Watanabe}}, ``{Heavy neutrino
  search in accelerator-based experiments},'' \bibinfo{journal}{JHEP}
  \textbf{\bibinfo{volume}{03}}, \bibinfo{pages}{125} (\bibinfo{year}{2013}),
  \eprint{1212.1062}.

\bibitem[{\citenamefont{Ballett et~al.}(2017)\citenamefont{Ballett, Pascoli,
  and Ross-Lonergan}}]{Ballett:2016opr}
\bibinfo{author}{\bibfnamefont{P.}~\bibnamefont{Ballett}},
  \bibinfo{author}{\bibfnamefont{S.}~\bibnamefont{Pascoli}}, \bibnamefont{and}
  \bibinfo{author}{\bibfnamefont{M.}~\bibnamefont{Ross-Lonergan}}, ``{MeV-scale
  sterile neutrino decays at the Fermilab Short-Baseline Neutrino program},''
  \bibinfo{journal}{JHEP} \textbf{\bibinfo{volume}{04}}, \bibinfo{pages}{102}
  (\bibinfo{year}{2017}), \eprint{1610.08512}.

\bibitem[{\citenamefont{Berryman et~al.}(2017)\citenamefont{Berryman,
  de~Gouv\^ea, Kelly, and Zhang}}]{Berryman:2017twh}
\bibinfo{author}{\bibfnamefont{J.~M.} \bibnamefont{Berryman}},
  \bibinfo{author}{\bibfnamefont{A.}~\bibnamefont{de~Gouv\^ea}},
  \bibinfo{author}{\bibfnamefont{K.~J.} \bibnamefont{Kelly}}, \bibnamefont{and}
  \bibinfo{author}{\bibfnamefont{Y.}~\bibnamefont{Zhang}}, ``{Dark Matter and
  Neutrino Mass from the Smallest Non-Abelian Chiral Dark Sector},''
  \bibinfo{journal}{Phys. Rev. D} \textbf{\bibinfo{volume}{96}},
  \bibinfo{pages}{075010} (\bibinfo{year}{2017}), \eprint{1706.02722}.

\bibitem[{\citenamefont{Coloma et~al.}(2017)\citenamefont{Coloma, Machado,
  Martinez-Soler, and Shoemaker}}]{Coloma:2017ppo}
\bibinfo{author}{\bibfnamefont{P.}~\bibnamefont{Coloma}},
  \bibinfo{author}{\bibfnamefont{P.~A.~N.} \bibnamefont{Machado}},
  \bibinfo{author}{\bibfnamefont{I.}~\bibnamefont{Martinez-Soler}},
  \bibnamefont{and} \bibinfo{author}{\bibfnamefont{I.~M.}
  \bibnamefont{Shoemaker}}, ``{Double-Cascade Events from New Physics in
  Icecube},'' \bibinfo{journal}{Phys. Rev. Lett.}
  \textbf{\bibinfo{volume}{119}}, \bibinfo{pages}{201804}
  (\bibinfo{year}{2017}), \eprint{1707.08573}.

\bibitem[{\citenamefont{Bondarenko et~al.}(2018)\citenamefont{Bondarenko,
  Boyarsky, Gorbunov, and Ruchayskiy}}]{Bondarenko:2018ptm}
\bibinfo{author}{\bibfnamefont{K.}~\bibnamefont{Bondarenko}},
  \bibinfo{author}{\bibfnamefont{A.}~\bibnamefont{Boyarsky}},
  \bibinfo{author}{\bibfnamefont{D.}~\bibnamefont{Gorbunov}}, \bibnamefont{and}
  \bibinfo{author}{\bibfnamefont{O.}~\bibnamefont{Ruchayskiy}},
  ``{Phenomenology of GeV-scale Heavy Neutral Leptons},''
  \bibinfo{journal}{JHEP} \textbf{\bibinfo{volume}{11}}, \bibinfo{pages}{032}
  (\bibinfo{year}{2018}), \eprint{1805.08567}.

\bibitem[{\citenamefont{Curtin et~al.}(2019)}]{Curtin:2018mvb}
\bibinfo{author}{\bibfnamefont{D.}~\bibnamefont{Curtin}} \bibnamefont{et~al.},
  ``{Long-Lived Particles at the Energy Frontier: The MATHUSLA Physics Case},''
  \bibinfo{journal}{Rept. Prog. Phys.} \textbf{\bibinfo{volume}{82}},
  \bibinfo{pages}{116201} (\bibinfo{year}{2019}), \eprint{1806.07396}.

\bibitem[{\citenamefont{Ahdida et~al.}(2019)}]{SHiP:2018xqw}
\bibinfo{author}{\bibfnamefont{C.}~\bibnamefont{Ahdida}} \bibnamefont{et~al.}
  (\bibinfo{collaboration}{SHiP}), ``{Sensitivity of the SHiP experiment to
  Heavy Neutral Leptons},'' \bibinfo{journal}{JHEP}
  \textbf{\bibinfo{volume}{04}}, \bibinfo{pages}{077} (\bibinfo{year}{2019}),
  \eprint{1811.00930}.

\bibitem[{\citenamefont{Ariga et~al.}(2019)}]{Ariga:2018uku}
\bibinfo{author}{\bibfnamefont{A.}~\bibnamefont{Ariga}} \bibnamefont{et~al.}
  (\bibinfo{collaboration}{FASER}), ``{FASER\textquoteright{}s physics reach
  for long-lived particles},'' \bibinfo{journal}{Phys. Rev. D}
  \textbf{\bibinfo{volume}{99}}, \bibinfo{pages}{095011}
  (\bibinfo{year}{2019}), \eprint{1811.12522}.

\bibitem[{\citenamefont{Krasnov}(2019)}]{Krasnov:2019kdc}
\bibinfo{author}{\bibfnamefont{I.}~\bibnamefont{Krasnov}}, ``{DUNE prospects in
  the search for sterile neutrinos},'' \bibinfo{journal}{Phys. Rev. D}
  \textbf{\bibinfo{volume}{100}}, \bibinfo{pages}{075023}
  (\bibinfo{year}{2019}), \eprint{1902.06099}.

\bibitem[{\citenamefont{Abe et~al.}(2019)}]{Abe:2019kgx}
\bibinfo{author}{\bibfnamefont{K.}~\bibnamefont{Abe}} \bibnamefont{et~al.}
  (\bibinfo{collaboration}{T2K}), ``{Search for heavy neutrinos with the T2K
  near detector ND280},'' \bibinfo{journal}{Phys. Rev. D}
  \textbf{\bibinfo{volume}{100}}, \bibinfo{pages}{052006}
  (\bibinfo{year}{2019}), \eprint{1902.07598}.

\bibitem[{\citenamefont{Ballett et~al.}(2020)\citenamefont{Ballett, Boschi, and
  Pascoli}}]{Ballett:2019bgd}
\bibinfo{author}{\bibfnamefont{P.}~\bibnamefont{Ballett}},
  \bibinfo{author}{\bibfnamefont{T.}~\bibnamefont{Boschi}}, \bibnamefont{and}
  \bibinfo{author}{\bibfnamefont{S.}~\bibnamefont{Pascoli}}, ``{Heavy Neutral
  Leptons from low-scale seesaws at the DUNE Near Detector},''
  \bibinfo{journal}{JHEP} \textbf{\bibinfo{volume}{20}}, \bibinfo{pages}{111}
  (\bibinfo{year}{2020}), \eprint{1905.00284}.

\bibitem[{\citenamefont{Drewes et~al.}(2019)\citenamefont{Drewes, Klari\'c, and
  Klose}}]{Drewes:2019byd}
\bibinfo{author}{\bibfnamefont{M.}~\bibnamefont{Drewes}},
  \bibinfo{author}{\bibfnamefont{J.}~\bibnamefont{Klari\'c}}, \bibnamefont{and}
  \bibinfo{author}{\bibfnamefont{P.}~\bibnamefont{Klose}}, ``{On Lepton Number
  Violation in Heavy Neutrino Decays at Colliders},'' \bibinfo{journal}{JHEP}
  \textbf{\bibinfo{volume}{11}}, \bibinfo{pages}{032} (\bibinfo{year}{2019}),
  \eprint{1907.13034}.

\bibitem[{\citenamefont{Arg\"uelles et~al.}(2020)\citenamefont{Arg\"uelles,
  Coloma, Hern\'andez, and Mu\~noz}}]{Arguelles:2019ziu}
\bibinfo{author}{\bibfnamefont{C.}~\bibnamefont{Arg\"uelles}},
  \bibinfo{author}{\bibfnamefont{P.}~\bibnamefont{Coloma}},
  \bibinfo{author}{\bibfnamefont{P.}~\bibnamefont{Hern\'andez}},
  \bibnamefont{and} \bibinfo{author}{\bibfnamefont{V.}~\bibnamefont{Mu\~noz}},
  ``{Searches for Atmospheric Long-Lived Particles},'' \bibinfo{journal}{JHEP}
  \textbf{\bibinfo{volume}{02}}, \bibinfo{pages}{190} (\bibinfo{year}{2020}),
  \eprint{1910.12839}.

\bibitem[{\citenamefont{Abratenko et~al.}(2020)}]{Abratenko:2019kez}
\bibinfo{author}{\bibfnamefont{P.}~\bibnamefont{Abratenko}}
  \bibnamefont{et~al.} (\bibinfo{collaboration}{MicroBooNE}), ``{Search for
  Heavy Neutral Leptons Decaying into Muon-Pion Pairs in the MicroBooNE
  Detector},'' \bibinfo{journal}{Phys. Rev. D} \textbf{\bibinfo{volume}{101}},
  \bibinfo{pages}{052001} (\bibinfo{year}{2020}), \eprint{1911.10545}.

\bibitem[{\citenamefont{Berryman et~al.}(2020)\citenamefont{Berryman,
  de~Gouv\^ea, Fox, Kayser, Kelly, and Raaf}}]{Berryman:2019dme}
\bibinfo{author}{\bibfnamefont{J.~M.} \bibnamefont{Berryman}},
  \bibinfo{author}{\bibfnamefont{A.}~\bibnamefont{de~Gouv\^ea}},
  \bibinfo{author}{\bibfnamefont{P.~J.} \bibnamefont{Fox}},
  \bibinfo{author}{\bibfnamefont{B.~J.} \bibnamefont{Kayser}},
  \bibinfo{author}{\bibfnamefont{K.~J.} \bibnamefont{Kelly}}, \bibnamefont{and}
  \bibinfo{author}{\bibfnamefont{J.~L.} \bibnamefont{Raaf}}, ``{Searches for
  Decays of New Particles in the DUNE Multi-Purpose Near Detector},''
  \bibinfo{journal}{JHEP} \textbf{\bibinfo{volume}{02}}, \bibinfo{pages}{174}
  (\bibinfo{year}{2020}), \eprint{1912.07622}.

\bibitem[{\citenamefont{Gorbunov et~al.}(2020)\citenamefont{Gorbunov, Krasnov,
  Kudenko, and Suvorov}}]{Gorbunov:2020rjx}
\bibinfo{author}{\bibfnamefont{D.}~\bibnamefont{Gorbunov}},
  \bibinfo{author}{\bibfnamefont{I.}~\bibnamefont{Krasnov}},
  \bibinfo{author}{\bibfnamefont{Y.}~\bibnamefont{Kudenko}}, \bibnamefont{and}
  \bibinfo{author}{\bibfnamefont{S.}~\bibnamefont{Suvorov}}, ``{Heavy Neutral
  Leptons from kaon decays in the SHiP experiment},'' \bibinfo{journal}{Phys.
  Lett. B} \textbf{\bibinfo{volume}{810}}, \bibinfo{pages}{135817}
  (\bibinfo{year}{2020}), \eprint{2004.07974}.

\bibitem[{\citenamefont{Coloma et~al.}(2021)\citenamefont{Coloma,
  Fern\'andez-Mart\'\i{}nez, Gonz\'alez-L\'opez, Hern\'andez-Garc\'\i{}a, and
  Pavlovic}}]{Coloma:2020lgy}
\bibinfo{author}{\bibfnamefont{P.}~\bibnamefont{Coloma}},
  \bibinfo{author}{\bibfnamefont{E.}~\bibnamefont{Fern\'andez-Mart\'\i{}nez}},
  \bibinfo{author}{\bibfnamefont{M.}~\bibnamefont{Gonz\'alez-L\'opez}},
  \bibinfo{author}{\bibfnamefont{J.}~\bibnamefont{Hern\'andez-Garc\'\i{}a}},
  \bibnamefont{and} \bibinfo{author}{\bibfnamefont{Z.}~\bibnamefont{Pavlovic}},
  ``{GeV-scale neutrinos: interactions with mesons and DUNE sensitivity},''
  \bibinfo{journal}{Eur. Phys. J. C} \textbf{\bibinfo{volume}{81}},
  \bibinfo{pages}{78} (\bibinfo{year}{2021}), \eprint{2007.03701}.

\bibitem[{\citenamefont{Batell et~al.}(2020)\citenamefont{Batell, Evans, Gori,
  and Rai}}]{Batell:2020vqn}
\bibinfo{author}{\bibfnamefont{B.}~\bibnamefont{Batell}},
  \bibinfo{author}{\bibfnamefont{J.~A.} \bibnamefont{Evans}},
  \bibinfo{author}{\bibfnamefont{S.}~\bibnamefont{Gori}}, \bibnamefont{and}
  \bibinfo{author}{\bibfnamefont{M.}~\bibnamefont{Rai}}, ``{Dark Scalars and
  Heavy Neutral Leptons at DarkQuest},''  (\bibinfo{year}{2020}),
  \eprint{2008.08108}.

\bibitem[{\citenamefont{De~Vries et~al.}(2021)\citenamefont{De~Vries, Dreiner,
  G\"unther, Wang, and Zhou}}]{deVries:2020qns}
\bibinfo{author}{\bibfnamefont{J.}~\bibnamefont{De~Vries}},
  \bibinfo{author}{\bibfnamefont{H.~K.} \bibnamefont{Dreiner}},
  \bibinfo{author}{\bibfnamefont{J.~Y.} \bibnamefont{G\"unther}},
  \bibinfo{author}{\bibfnamefont{Z.~S.} \bibnamefont{Wang}}, \bibnamefont{and}
  \bibinfo{author}{\bibfnamefont{G.}~\bibnamefont{Zhou}}, ``{Long-lived Sterile
  Neutrinos at the LHC in Effective Field Theory},'' \bibinfo{journal}{JHEP}
  \textbf{\bibinfo{volume}{03}}, \bibinfo{pages}{148} (\bibinfo{year}{2021}),
  \eprint{2010.07305}.

\bibitem[{\citenamefont{Plestid}(2020)}]{Plestid:2020ssy}
\bibinfo{author}{\bibfnamefont{R.}~\bibnamefont{Plestid}}, ``{Luminous solar
  neutrinos II: Mass-mixing portals},''  (\bibinfo{year}{2020}),
  \eprint{2010.09523}.

\bibitem[{\citenamefont{Breitbach et~al.}(2021)\citenamefont{Breitbach,
  Buonocore, Frugiuele, Kopp, and Mittnacht}}]{Breitbach:2021gvv}
\bibinfo{author}{\bibfnamefont{M.}~\bibnamefont{Breitbach}},
  \bibinfo{author}{\bibfnamefont{L.}~\bibnamefont{Buonocore}},
  \bibinfo{author}{\bibfnamefont{C.}~\bibnamefont{Frugiuele}},
  \bibinfo{author}{\bibfnamefont{J.}~\bibnamefont{Kopp}}, \bibnamefont{and}
  \bibinfo{author}{\bibfnamefont{L.}~\bibnamefont{Mittnacht}}, ``{Searching for
  Physics Beyond the Standard Model in an Off-Axis DUNE Near Detector},''
  (\bibinfo{year}{2021}), \eprint{2102.03383}.

\end{thebibliography}

\end{document}